\documentclass[traditabstract]{aa} % for the abstract without structuration 
\usepackage{graphicx}
\usepackage{txfonts}
\usepackage{booktabs}
\usepackage{multirow}
\usepackage{natbib}
\usepackage{wasysym} 
\usepackage{wrapfig,lipsum,booktabs}

\bibpunct{(}{)}{;}{a}{}{,}
\newcommand{\teff}{\mbox{$\rm T_{\rm eff}$}}
\newcommand{\logg}{\mbox{$\log g$}}
\newcommand{\vsini}{\mbox{$v \sin i^{\star}$}}
\newcommand{\mictrb}{\mbox{$\xi_{\rm t}$}}
\newcommand{\mactrb}{\mbox{$v_{\rm mac}$}}

\newcommand{\kms}{\mbox{km\,s$^{-1}$}}
\newcommand{\halpha}{\mbox{$H_\alpha$}}
\newcommand{\gammaS}{\mbox{$\gamma_{\rm SOPHIE}$}}
\newcommand{\gammaC}{\mbox{$\gamma_{\rm CORALIE}$}}

\newcommand{\rhostar}{\ensuremath{\rho_\star}}
\newcommand{\rhosun}{\ensuremath{\rho_\odot}}
\newcommand{\rhoj}{\ensuremath{\rho_{\rm J}}}
\newcommand{\rhopl}{\ensuremath{\rho_{\rm pl}}}
\newcommand{\rj}{R\ensuremath{_{\rm J}}}
\newcommand{\mj}{M\ensuremath{_{\rm J}}}
\newcommand{\rsun}{R\ensuremath{_\odot}}
\newcommand{\msun}{M\ensuremath{_\odot}}
\newcommand{\rpl}{\ensuremath{R_{\rm pl}}}
\newcommand{\mpl}{\ensuremath{M_{\rm pl}}}
\newcommand{\rstar}{\ensuremath{R_\star}}
\newcommand{\mstar}{\ensuremath{M_\star}}
\newcommand{\degree}{\mbox{\ensuremath{^\circ}}}   %fractional degree symbol

\def\ecos{$e \cos \omega$}
\def\esin{$e \sin \omega$}
\def\secos{$\sqrt{e} \cos \omega$}
\def\sesin{$\sqrt{e} \sin \omega$}
\def\feh{[Fe/H]}

\begin{document}

\title{WASP-54b, WASP-56b and WASP-57b:\\
  Three new sub-Jupiter mass planets from
  SuperWASP \thanks{Spectroscopic and photometric data are 
    available in electronic form at the CDS via anonymous ftp to cdsarc.u-strasbg.fr
    (130.79.128.5) or via
    http://cdsweb.u-strasbg.fr/cgi-bin/qcat?J/A+A/}}

\author{F. Faedi$^*$ \inst{1,2} \and D. Pollacco$^*$ \inst{1,2} \and
  S. C. C. Barros \inst{3} \and D. Brown \inst{4} \and A. Collier
  Cameron \inst{4} \and A.P. Doyle \inst{5} \and R. Enoch \inst{4} \and M. Gillon \inst{6}
  \and Y. G\'omez Maqueo Chew$^*$ \inst{1,2,7} \and G. H\'ebrard
  \inst{8,9} \and M. Lendl \inst{10} \and C. Liebig \inst{4} \and
  B. Smalley \inst{5} \and A. H. M. J. Triaud \inst{10} \and
  R. G. West \inst{11} \and P. J. Wheatley \inst{1} \and K. A. Alsubai
  \inst{12} \and D. R. Anderson \inst{5} \and D. Armstrong$^*$
  \inst{1,2} \and J. Bento \inst{1} \and J. Bochinski \inst{13} \and F. Bouchy \inst{8,9}
  \and R. Busuttil \inst{13} \and L. Fossati \inst{14} \and A. Fumel
  \inst{6} \and C. A. Haswell \inst{13} \and C. Hellier \inst{5} \and
  S. Holmes \inst{13} \and E. Jehin \inst{6} \and U. Kolb \inst{13}
  \and J. McCormac$^*$ \inst{15,2,1} \and G. R. M. Miller \inst{4} \and
  C. Moutou \inst{3} \and A. J. Norton \inst{13} \and N. Parley
  \inst{4} \and D. Queloz \inst{10} \and A. Santerne \inst{3} \and I. Skillen \inst{15} \and
  A. M. S. Smith \inst{5} \and S. Udry \inst{10} \and C. Watson
  \inst{2} }

\institute{Department of Physics, University of Warwick, Coventry CV4 7AL, UK \\
  \email{f.faedi@warwick.ac.uk\\
  $^*$ part of the work was carried out while at Queen's University Belfast }
  \and Astrophysics Research Centre, School of Mathematics and Physics, Queen's University Belfast, University Road, Belfast BT7 1NN.\\
  \and Aix Marseille Université, CNRS, LAM (Laboratoire d'Astrophysique de Marseille) UMR 7326, 13388, Marseille, France\\
  \and School of Physics and Astronomy,  University of St. Andrews, St. Andrews, Fife KY16 9SS, UK \\
  \and Astrophysics Group, Keele University,  Staffordshire, ST5 5BG, UK \\
  \and Universit\'e de Li\`ege, All\'ee du 6 ao$\hat {\rm u}$t 17, Sart Tilman, Li\`ege 1, Belgium\\
  \and Physics and Astronomy Department, Vanderbilt University, Nashville, Tennessee, USA\\
  \and Institut d'Astrophysique de Paris, UMR7095 CNRS, Universit\'e Pierre \& Marie Curie, France \\
  \and Observatoire de Haute-Provence, CNRS/OAMP, 04870 St Michel l'Observatoire, France \\
  \and Observatoire astronomique de l'Universit\'e de Gen\`eve, 51 ch.\ des Maillettes, 1290 Sauverny, Switzerland \\
  \and Department of Physics and Astronomy, University of Leicester, Leicester, LE1 7RH \\
  \and Qatar Foundation, P.O.BOX 5825, Doha, Qatar\\
  \and Department of Physical Sciences, The Open University, Milton Keynes, MK7 6AA, UK \\
  \and Argelander-Institut f\"{u}r Astronomie der Universit\"{a}t Bonn, Auf dem Hügel 71, 53121, Bonn, Germany \\
  \and Isaac Newton Group of Telescopes, Apartado de Correos 321, E-38700 Santa Cruz de Palma, Spain\\
}

   \date{Received; accepted }

   \abstract{We present three newly discovered sub-Jupiter mass planets
     from the SuperWASP survey: WASP-54b is a heavily bloated planet
     of mass 0.636$^{+0.025}_{-0.024}$ \mj\ and radius
     1.653$^{+0.090}_{-0.083}$ \rj. It orbits a F9 star, evolving off
     the main sequence, every 3.69 days. Our MCMC fit of the system
     yields a slightly eccentric orbit ($e=0.067^{+0.033}_{-0.025}$)
     for WASP-54b. We investigated further the veracity of our
     detection of the eccentric orbit for WASP-54b, and we find that
     it could be real. However, given the brightness of WASP-54
     V=10.42 magnitudes, we encourage observations of a secondary
     eclipse to draw robust conclusions on both the orbital
     eccentricity and the thermal structure of the planet. WASP-56b
     and WASP-57b have masses of 0.571$^{+0.034}_{-0.035}$ \mj\ and
     $0.672^{+0.049}_{-0.046}$ \mj~, respectively; and radii of
     $1.092^{+0.035}_{-0.033}$ \rj\ for WASP-56b and
     $0.916^{+0.017}_{-0.014}$ \rj\ for WASP-57b. They orbit main
     sequence stars of spectral type G6 every 4.67 and 2.84 days,
     respectively. WASP-56b and WASP-57b show no radius anomaly and a
     high density possibly implying a large core of heavy elements;
     possibly as high as $\sim$50 M$_{\oplus}$ in the case of
     WASP-57b. However, the composition of the deep interior of
     exoplanets remain still undetermined. Thus, more exoplanet
     discoveries such as the ones presented in this paper, are needed
     to understand and constrain giant planets' physical properties.

   }


    \titlerunning{Three new exoplanets from WASP}
    \authorrunning{Faedi F.}
    \maketitle
%
%________________________________________________________________

\section{Introduction}

To date the number of extrasolar planets for which precise
measurements of masses and radii are available amounts to more than a
hundred.  Although these systems are mostly Jupiter--like gas giants
they have revealed an extraordinary variety of physical and dynamical
properties that have had a profound impact on our knowledge of
planetary structure, formation and evolution and unveiled the
complexity of these processes \citep[see][and references there
in]{Baraffe2010}. Transit surveys such as SuperWASP
\citep{Pollacco2006} have been extremely successful in providing great
insight into the properties of extrasolar planets and their host stars
(see e.g., \citealt{Baraffe2010}). Ground-based surveys excel in
discovering systems with peculiar/exotic characteristics. Subtle
differences in their observing strategies can yield unexpected
selection effects impacting the emerging distributions of planetary
and stellar properties such as orbital periods, planetary radii and
stellar metallicity (see e.g., \citealt{Cameron2011} for a
discussion). For example WASP-17b \citep{Anderson2010a} is a highly
inflated ($\rpl=1.99\rj$), very low density planet in a
tilted/retrograde orbit, HAT-P-32b \citep{Hartman2011} could be close
to filling its Roche Lobe $\rpl=2.05\rj$ (if the best fit eccentric
orbit is adopted\footnote{However, we stress here that the favoured
  circular solution results in a best fit radius of
  $1.789\pm0.025\rj$}), thus possibly losing its gaseous envelope, and
the heavily irradiated and bloated WASP-12b (\citealt{Hebb2009}), has
a Carbon rich atmosphere (\citealt{Kopparapu2012};
\citealt{Fossati2010}), and is undergoing atmospheric evaporation
(\citealt{Llama2011}; \citealt{LecavelierDesEtangs2010}) losing mass
to its host star at a rate $\sim$10$^{-7}$~\mj~yr$^{-1}$
\citep{Li2010}. On the opposite side of the spectrum of planetary
parameters, the highly dense Saturn-mass planet HD\,149026b is thought
to have a core of heavy elements with $\sim$70M$_{\oplus}$, needed to
explain its small radius (e.g., \citealt{Sato2005}, and
\citealt{Carter2009}), and the massive WASP-18b (\mpl = 10\mj,
\citealt{Hellier2009}), is in an orbit so close to its host star with
period of $\sim$0.94~d and eccentricity $e = 0.02$, that it might
induce significant tidal effects probably spinning up its host star
\citep{Brown2011}. Observations revealed that some planets are larger
than expected from standard coreless models (e.g.,
\citealt{Fortney2007}, \citealt{Baraffe2008}) and that the planetary
radius is correlated with the planet equilibrium temperature and
anti-correlated with stellar metallicity (see \citealt{Guillot2006};
\citealt{Laughlin2011}; \citealt{Enoch2010}; \citealt{Faedi2011}). For
these systems different theoretical explanations have been proposed
for example, tidal heating due to unseen companions pumping up the
eccentricity (\citealt{Bodenheimer2001}; and
\citealt{Bodenheimer2003}), kinetic heating due to the breaking of
atmospheric waves \citep{Guillot2002}, enhanced atmospheric opacity
\citep{Burrows2007}, semi convection \citep{Chabrier2007}, and finally
ohmic heating (\citealt[2010]{Batygin2011}; and
\citealt{Perna2012}). While each individual mechanism would presumably
affect all hot Jupiters to some extent, they can not explain the
entirety of the observed radii (\citealt{Fortney2010};
\citealt{Leconte2010}; \citealt{Perna2012}).  More complex thermal
evolution models are necessary to fully understand their cooling
history.

Recently, the Kepler satellite mission released a large number of
planet candidates ($>$2000) and showed that Neptune-size candidates
and Super-Earths ($>$ 76\% of Kepler planet candidates) are common
around solar-type stars (e.g., \citealt{Borucki2011}, and
\citealt{Batalha2012}). Although these discoveries are fundamental for
a statistically significant study of planetary populations and
structure in the low--mass regime, the majority of these candidates
orbit stars that are intrinsically faint (V $>$ 13.5 for $\sim$78\% of
the sample of \citealt{Borucki2011}) compared to those observed from
ground-based transit surveys, making exoplanet confirmation and
characterisation extremely challenging if not impossible. Thus, more
bright examples of transiting planets are needed to extend the
currently known parameter space in order to provide observation
constraints to test theoretical models of exoplanet structure,
formation and evolution. Additionally, bright gas giant planets also
allow study of their atmospheres via transmission and emission
spectroscopy, and thus provide interesting candidates for future
characterisation studies from the ground (e.g. VLT and e-ELT) and from
space (e.g. PLATO, JWST, EChO, and FINESSE).

Here we describe the properties of three newly discovered transiting
exoplanets from the WASP survey: WASP-54b, WASP-56b, and WASP-57b. The
paper is structured as follows: in $\S$~2 we describe the
observations, including the WASP discovery data and follow up
photometric and spectroscopic observations which establish the
planetary nature of the transiting objects. In $\S$~3 we present our
results for the derived system parameters for the three planets, as
well as the individual stellar and planetary properties. Finally, in
$\S$~4 we discuss the implication of these discoveries, their physical
properties and how they add information to the currently explored
mass-radius parameter space.

%__________________________________________________________________

\section{Observations}

The stars 1SWASP J134149.02-000741.0 (2MASS~J13414903-0007410)
hereafter WASP-54; 1SWASP J121327.90+230320.2
(2MASS~J12132790+2303205) hereafter WASP-56; and 1SWASP
J145516.84-020327.5 (2MASS~J14551682-0203275) hereafter WASP-57; have
been identified in several northern sky catalogues which provide
broad-band optical \citep{Nomadcat} and infra-red 2MASS magnitudes
\citep{2MASS} as well as proper motion information. Coordinates,
broad-band magnitudes and proper motion of the stars are from the
NOMAD 1.0 catalogue and are given in Table~\ref{table1}.

\subsection{SuperWASP observations}

The WASP North and South telescopes are located in La Palma (ORM -
Canary Islands) and Sutherland (SAAO - South Africa), respectively.
Each telescope consists of 8 Canon 200mm f/1.8 focal lenses coupled to
e2v $2048\times2048$ pixel CCDs, which yield a field of view of
$7.8\times7.8$ square degrees, and a pixel scale of 13.7\arcsec~
\citep{Pollacco2006}.\\

WASP-56 ($V=11.5$) is located in the northern hemisphere with
Declination $\delta \sim +23h$ and thus it is only observed by the
SuperWASP-North telescope; WASP-54 and WASP-57 ($V=10.42$ and
$V=13.04$, respectively) are located in an equatorial region of sky
monitored by both WASP instruments, however only WASP-54 has been
observed simultaneously by both telescopes, with a significantly
increased observing coverage on the target. In January 2009 the
SuperWASP-N telescope underwent a system upgrade that improved our
control over the main sources of red noise, such as
temperature-dependent focus changes (\citealt{Barros2011};
\citealt{Faedi2011}). This upgrade resulted in better quality data
and increased the number of planet detections.

All WASP data for the three new planet-hosting stars were processed
with the custom-built reduction pipeline described in
\citet{Pollacco2006}. The resulting light curves were analysed using
our implementation of the Box Least-Squares fitting and SysRem de-trending
algorithms (see \citealt{Cameron2006}; \citealt{Kovacs2002};
\citealt{Tamuz2005}), to search for signatures of planetary transits.
Once the candidate planets were flagged, a series of multi-season,
multi-camera analyses were performed to strengthen the candidate
detection. In addition different de-trending algorithms (e.g., TFA,
\citealt{Kovacs2005}) were used on one season and multi-season light
curves to confirm the transit signal and the physical parameters of
the planet candidate. These additional tests allow a more thorough
analysis of the stellar and planetary parameters derived solely from
the WASP data thus helping in the identification of the best
candidates, as well as to reject possible spurious detections.\\

\begin{figure} 
\centering
  \includegraphics[width=0.5\textwidth]{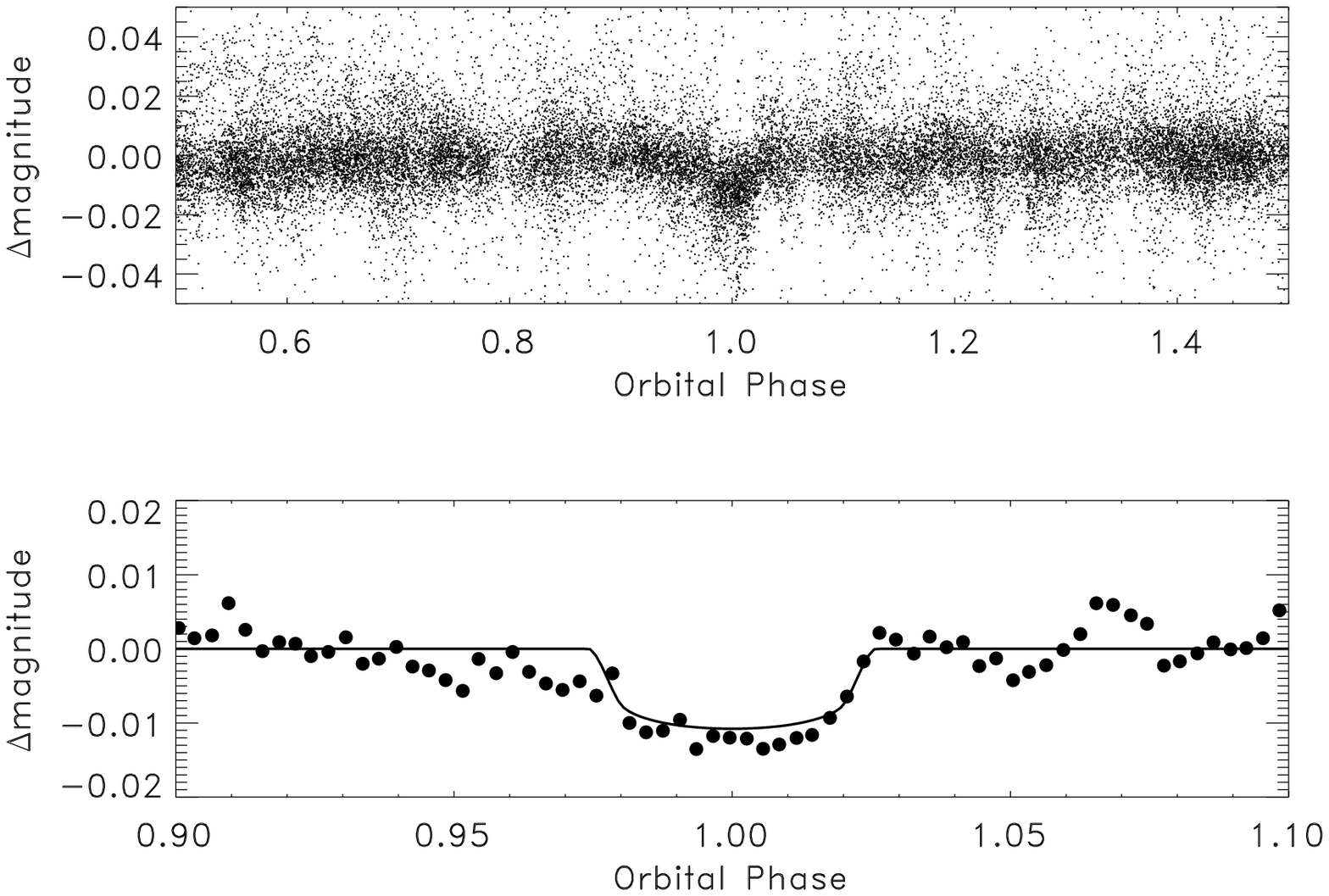}
 \caption{ {\em Upper panel}: Discovery light curve of WASP-54b phase folded
  on the ephemeris given in Table~\ref{W54_params}.
  {\em Lower panel}: binned WASP-54b light curve. Black-solid line, is the best-fit
  transit model estimated using the formalism from \citet{Mandel2002}.
} \label{W54_waspLC}
\end{figure}

- {\bf WASP-54} was first observed in 2008, February 19. The same
field was observed again in 2009, 2010 and 2011 by both WASP
telescopes. This resulted in a total of 29938 photometric data points,
of which 1661 are during transit. A total of 58 partial or full
transits were observed with an improvement in $\chi^2$ of the
box-shaped model over the flat light curve of $\Delta\chi^2=-701$, and
signal-to-red noise value \citep{Cameron2006} of $SN_{red} = -13.02$.
When combined, the WASP data of WASP-54, showed a characteristic
periodic dip with a period of $P=3.69$~days, duration
$T_{14}\sim270$~mins, and a depth $\sim11.5$~mmag.
igure~\ref{W54_waspLC} shows the discovery photometry of WASP-54b
phase folded on the period above, and the binned phased light curve.\\

\begin{figure}
\centering
\includegraphics[width=0.5\textwidth]{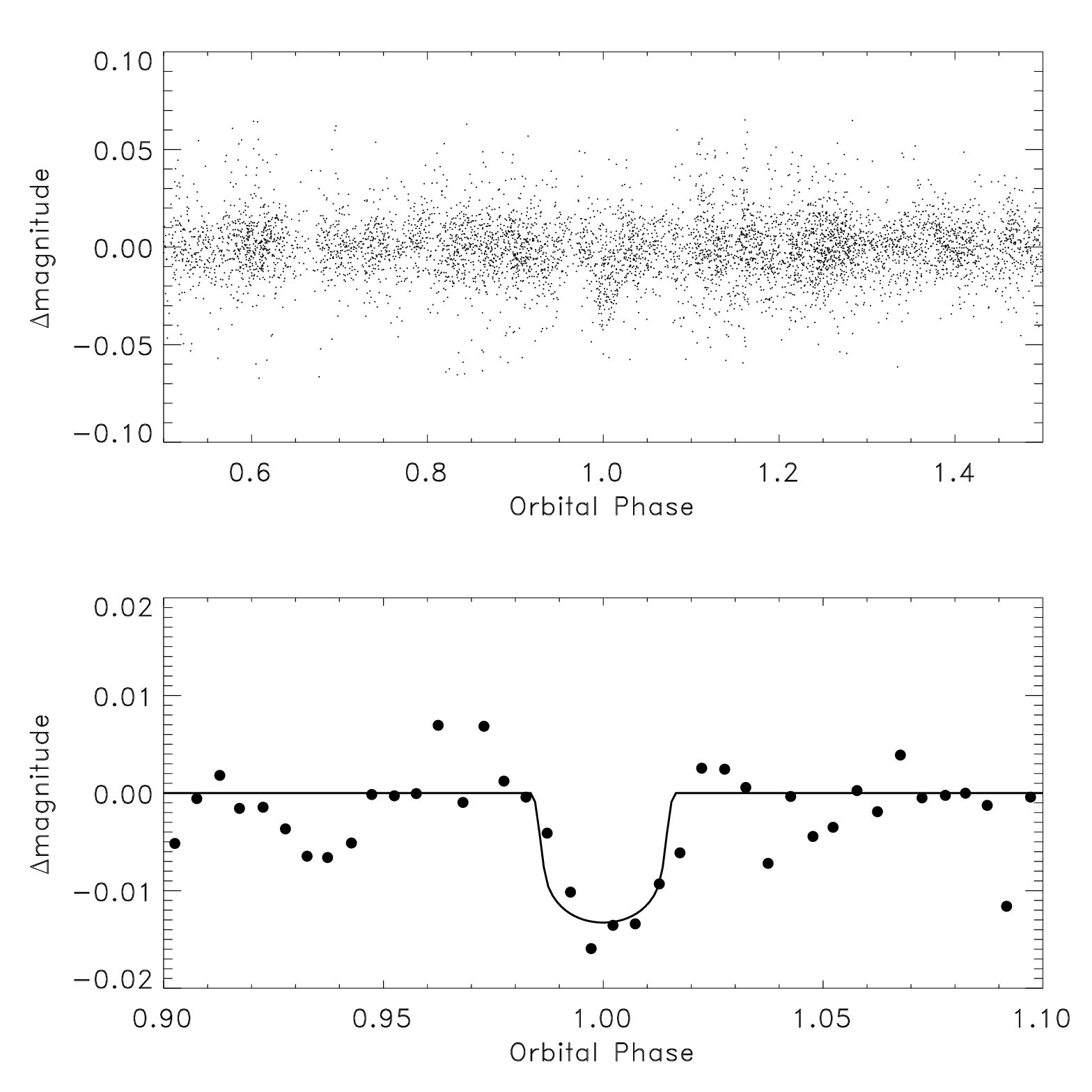}
\caption{ {\em Upper panel}: Discovery light curve of WASP-56b phase
  folded on the ephemeris given in Table~\ref{W56_params}. {\em Lower
    panel}: binned WASP-56b light curve. Black-solid line, is the
  best-fit transit model estimated using the formalism from
  \citet{Mandel2002}. } 
\label{W56_waspLC} 
\end{figure}

- {\bf WASP-56} was first observed during our pilot survey in May 2004
by SuperWASP-North. The same field was also observed in 2006 and 2007
yielding a total of 16441 individual photometric observations.
SuperWASP first began operating in the northern hemisphere in 2004,
observing in white light with the spectral transmission defined by the
optics, detectors, and atmosphere. During the 2004 season the phase
coverage for WASP-56b was too sparse to yield a robust detection with
only $\sim 10$ points falling during the transit phase. Later in 2006
a broad-band filter (400 -- 700~nm) was introduced and with more data
available multi-season runs confirmed the transit detection. Over the
three seasons a total of 14 partial or full transits were observed,
yielding 300 observations in transit, with a $\Delta\chi^2=-213$
improvement over the flat light curve, and $SN_{red} = -7.02$. The
combined WASP light curves, plotted in Figure \ref{W56_waspLC}, show
the detected transit signal of period = 4.61~days, depth =
$\sim13$~mmag, and duration $T_{14}\sim214$~mins.\\

\begin{figure} \centering
  \includegraphics[width=0.5\textwidth]{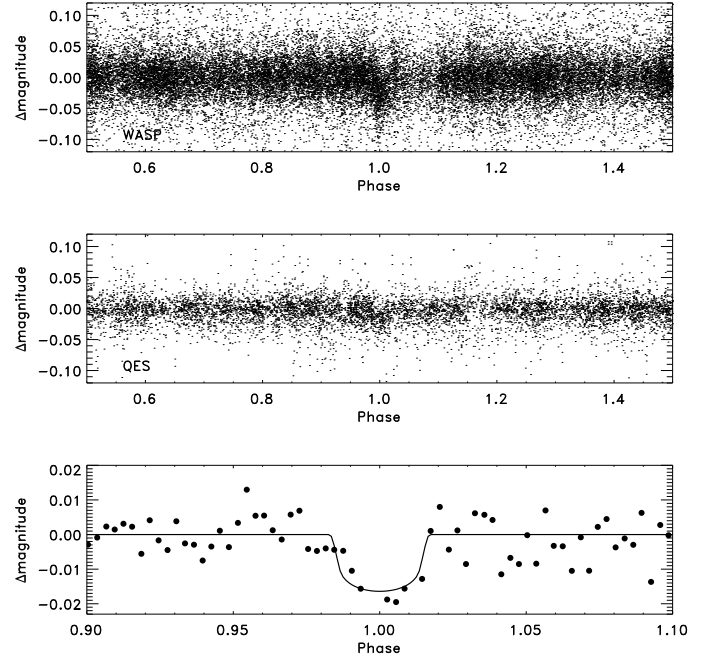}
  \caption{ {\em Upper panel}: Discovery WASP light curve of WASP-57b
    phase folded on the ephemeris given in Table~\ref{W56_params}. {\em Middle panel}: QES light curve of WASP-57b.
  {\em Lower panel}: binned WASP light curve of WASP-57b. Black-solid line, is the best-fit
  transit model estimated using the formalism from
  \citet{Mandel2002}.} \label{W57_waspLC} 
\end{figure}

- {\bf WASP-57b} was first observed in March 2008 and subsequently in
Spring 2010. A total of 30172 points were taken of which about 855
were during transit. About 65 full or partial transits were observed
overall with a $\Delta\chi^2=-151$, and $SN_{red} = -6.20$. Figure
\ref{W57_waspLC}-{\em upper panel} shows the combined WASP light
curves folded on the detected orbital period of 2.84~days.
Additionally, for WASP-57b there is photometric coverage from the
Qatar Exoplanet Survey (QES, \citealt{Alsubai2011}) and the phase
folded QES light curve is shown in Figure \ref{W57_waspLC}-{\em middle
  panel}. In both WASP and QES light curves the transit signal was
identified with a period $\sim$2.84~days, duration
$T_{14}\sim138$~mins, and transit depth of $\sim 17$~mmag.

\begin{table} 
\caption[]{Photometric properties of the stars
    WASP-54, WASP-56 and WASP-57. The broad-band magnitudes and proper
    motion are obtained from the NOMAD~1.0 catalogue.}
\label{table1}
%\begin{center}
\begin{tabular}{lccc}
\hline
\hline
 Parameter    & WASP-54 & WASP-56 & WASP-57 \\
 \hline
 ${\rm RA (J2000)}$		&13:41:49.02		&12:13:27.90	 	&14:55:16.84\\ 
${\rm Dec (J2000)}$		&$-$00:07:41.0	&+23:03:20.2		&$-$02:03:27.5\\
 %                    			&				&&				&&\\
${\rm B}$				&$10.98\pm0.07$   &$12.74\pm0.28$	        &$13.6\pm0.5$\\
${\rm V}$			        &$10.42\pm0.06$	&$11.484\pm0.115$	&$13.04\pm0.25$\\
${\rm R}$				&$10.0\pm0.3$	&$10.7\pm0.3$	  	&$12.7\pm0.3$\\	
${\rm I}$				&$9.773\pm0.053$ &$11.388\pm0.087$	&$12.243\pm0.107$\\
${\rm J}$				&$9.365\pm0.022$ &$10.874\pm0.021$	&$11.625\pm0.024$\\
${\rm H}$			&$9.135\pm0.027$ &$10.603\pm0.022$	&$11.292\pm0.024$\\
${\rm K}$				&$9.035\pm0.023$ &$10.532\pm0.019$	&$11.244\pm0.026$\\
$\mu_{\alpha}$ (mas/yr)	&$-9.8\pm1.3$       &$-34.9\pm0.8$&$-22.0\pm5.4$\\
$\mu_{\delta}$   (mas/yr)	&$-23.5\pm1.2$	&$2.9\pm0.7$		&$-0.6\pm5.4$\\
\hline\\
\end{tabular}
%\end{center}
\end{table}
%______________________________________________ 
 
\subsection{Low S/N photometry}
Several observing facilities are available to the WASP consortium and
are generally used to obtain multi-band low-resolution photometry to
confirm the presence of the transit signal detected in the WASP light
curves. This is particularly useful in case of unreliable ephemerides,
and in case the transit period is such that follow up from a
particular site is more challenging. Small-to-medium sized telescopes
such as the remote-controlled 17-inch PIRATE telescope in the
Observatori Astronomic de Mallorca \citep{Holmes2011}, together with
the James Gregory 0.94~m telescope (JGT) at the University of
St. Andrews, provide higher precision, higher spatial resolution
photometry as compared to WASP, and thus have an important role as a
link in the planet-finding chain, reducing the amount of large
telescope time spent on false-positives. Observations of WASP-56 were
obtained with both PIRATE and JGT, while observations of WASP-54 were
obtained only with PIRATE.

Multiple Markov-Chain Monte Carlo (MCMC) chains have been obtained for
both systems to assess the significance of adding the PIRATE and JGT
light curves to the corresponding dataset in determining the transit
model, in particular the impact parameter, the transit duration, and
$a/\rstar$. We conclude that for WASP-54 the effect is not
significant, never the less, the PIRATE light curves were included in
our final analysis presented in section \S \ref{results}. In the case
of WASP-56 instead, because we only have a partial TRAPPIST light
curve (see section \S 2.4), the full JGT light curve,
although of lower quality, is crucial to better constrain the transit
ingress/egress time, impact parameter and $a/\rstar$, allowing us to
relax the main sequence mass-radius constraint.

\subsection{Spectroscopic follow up}

WASP-54, 56 and 57 were observed during our follow up campaign in
Spring 2011 with the SOPHIE spectrograph mounted at the 1.93~m
telescope (\citealt{Perruchot2008}; \citealt{Bouchy2009}) at
Observatoire de Haute-Provence (OHP), and the CORALIE spectrograph
mounted at the 1.2~m Euler-Swiss telescope at La Silla, Chile
(\citealt{Baranne1996}; \citealt{Queloz2000}; \citealt{Pepe2002}). We
used SOPHIE in high efficiency mode (R = 40\,000) and obtained
observations with very similar signal-to-noise ratio ($\sim$30), in
order to minimise systematic errors (e.g., the Charge Transfer
Inefficiency effect of the CCD, \citealt{Bouchy2009}). Wavelength
calibration with a Thorium-Argon lamp was performed every $\sim$2
hours, allowing the interpolation of the spectral drift of SOPHIE
($<3$~m\,s$^{-1}$ per hour; see \citealt{Boisse2010}). Two 3$\arcsec$
diameter optical fibers were used; the first centered on the target
and the second on the sky to simultaneously measure the background to
remove contamination from scattered moonlight. During SOPHIE
observations of WASP-54, 56 and 57 the contribution from scattered
moonlight was negligible as it was well shifted from the targets'
radial velocities. The CORALIE observations of WASP-54 and WASP-57
were obtained during dark/grey time to minimise moonlight
contamination. The data were processed with the SOPHIE and CORALIE
standard data reduction pipelines, respectively. The radial velocity
uncertainties were evaluated including known systematics such as
guiding and centering errors \citep{Boisse2010}, and wavelength
calibration uncertainties. All spectra were single-lined.

For each planetary system the radial velocities were computed from a
weighted cross-correlation of each spectrum with a numerical mask of
spectral type G2, as described in \citet{Baranne1996} and
\citet{Pepe2002}. To test for possible stellar impostors we performed
the cross-correlation with masks of different stellar spectral types
(e.g. F0, K5 and M5). For each mask we obtained similar radial
velocity variations, thus rejecting a blended eclipsing system of
stars with unequal masses as a possible cause of the variation.

We present in Tables \ref{W54_RVtable}, \ref{W56_RVtable}, and
\ref{W57_RVtable} the spectroscopic measurements of WASP-54, 56 and 57
together with their line bisectors (V$_{span}$). In each Table we list
the Barycentric Julian date (BJD), the stellar radial velocities
(RVs), their uncertainties, the bisector span measurements, and the
instrument used. In column 6, we list the radial velocity measurements
after subtracting the zero point offset to CORALIE and SOPHIE data
respectively (the zero-point offsets are listed in
Table~\ref{W54_params}, and Table~\ref{W56_params} respectively). In
column 7 we also give the line bisectors after subtracting the mean
value for SOPHIE and CORALIE respectively, and finally, in column 8,
the radial velocity residuals to the best-fit Keplerian model. The
Root-Mean-Square ($RMS$) of the residuals to the best-fit Keplerian
models are as follow: $RMS = 18.9$~ms$^{-1}$ for WASP-54, $RMS =
19.5$~ms$^{-1}$ for WASP-56, and $RMS = 24.4$ms$^{-1}$ for WASP-57.

%%%%%%%%%%%%%%%%%%%%%% 
%% RV data table 
%%%%%%%%%%%%%%%%%%%%%% 

%\twocolumn 

\begin{figure}
   \centering
   \includegraphics[width=0.5\textwidth]{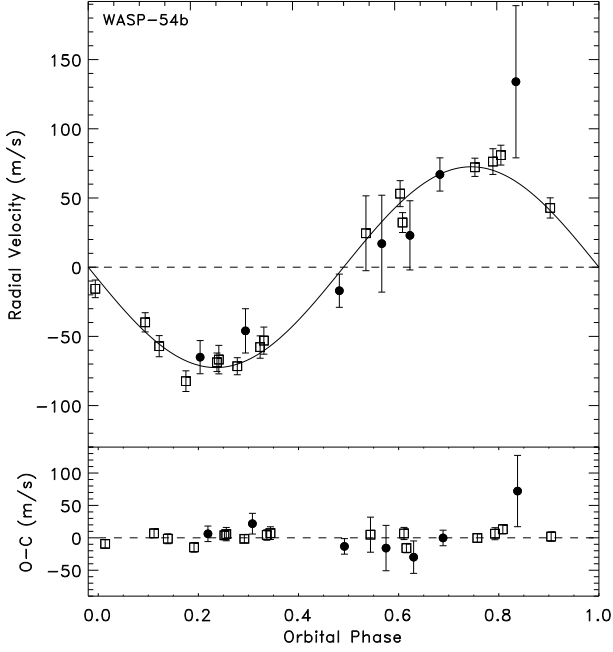}
   \caption{{\em Upper panel}: Phase folded radial velocity
     measurements of WASP-54 obtained combining data from SOPHIE
     (filled-circles) and CORALIE (open-squares) spectrographs.
     Superimposed is the best-fit model RV curve with parameters from
     Table~\ref{W54_params}. The centre-of-mass velocity for each data
     set was subtracted from the RVs (\gammaS = -3.1109~\kms~and
     \gammaC = -3.1335~\kms). {\em Lower panel}: Residuals from the RV
     orbital fit plotted against time.}
    \label{W54_RVplot}%
   \end{figure}

   \begin{figure}
  \centering
   \includegraphics[width=0.5\textwidth]{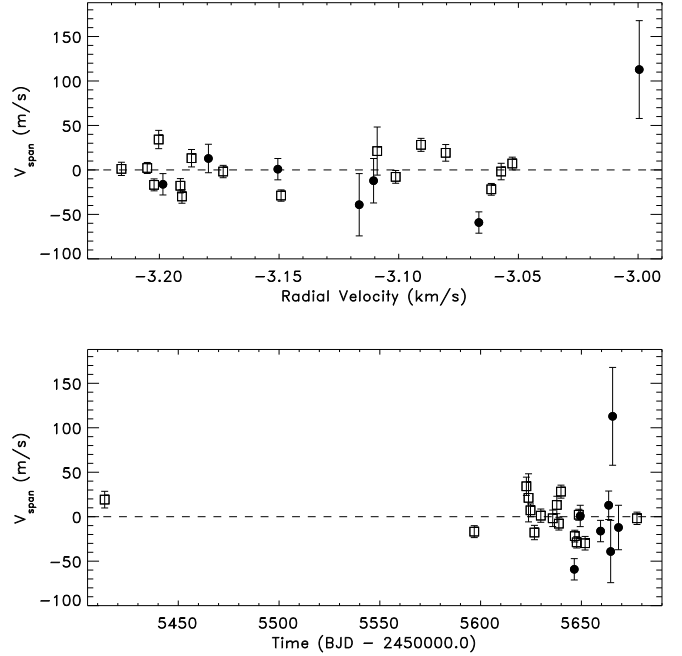}
   \caption{{\em Upper panel}: The bisector span measurements
        of WASP-54 as a function of radial velocity, values are
        shifted to a zero-mean ($<$$V_{span}$$>_{SOPHIE} =
        -29$~m\,s$^{-1}$, $<$$V_{span}$$>_{CORALIE} =
        48$~m\,s$^{-1}$). {\em Lower panel}: The bisector span
        measurements as a function of time (BJD\,--\,245\,0000.0). The
        bisector span shows no significant variation nor correlation
        with the RVs, suggesting that the signal is mainly due to
        Doppler shifts of the stellar lines rather than stellar
        profile variations due to stellar activity or a blended
        eclipsing binary.} \label{W54_Vspan}
    \end{figure}

\begin{figure}
  \centering
   \includegraphics[width=0.5\textwidth]{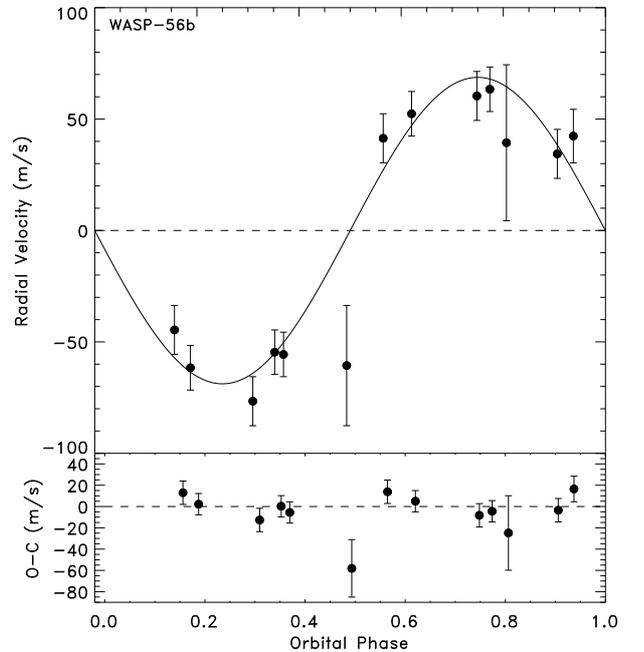}
      \caption{{\em Upper panel}: Similar to Figure \ref{W54_RVplot}, the phase folded radial velocity measurements of WASP-56. 
        The centre-of-mass velocity for the SOPHIE data was subtracted
        from the RVs (\gammaS$ = -4.6816$~\kms). {\em Lower panel}:
        Residuals from the RV orbital fit plotted against time.}
\label{W56_RVplot}%
\end{figure}

\begin{figure}
   \centering
   \includegraphics[width=0.5\textwidth]{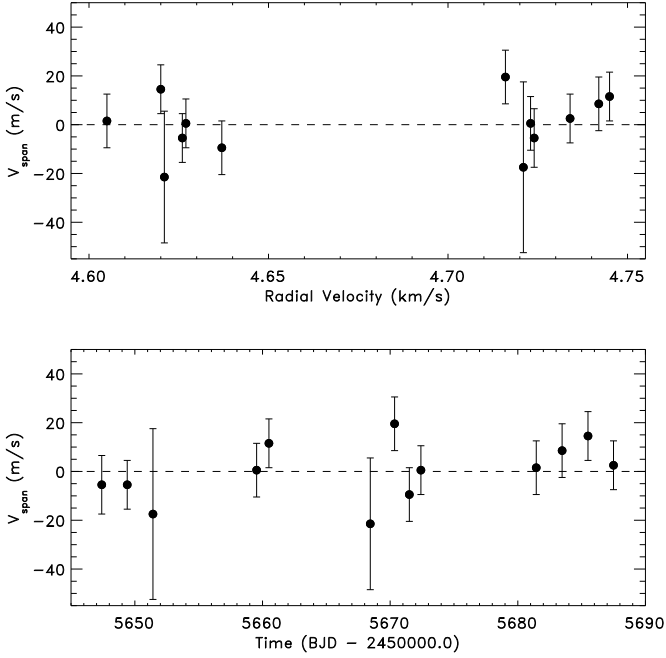}
      \caption{
        {\em Upper panel}: The bisector span measurements for
        WASP-56 as a function of radial velocity, values are shifted
        to a zero-mean ($<$$V_{span}$$>_{SOPHIE} = -40$~m\,s$^{-1}$).
        {\em Lower panel}: The bisector span measurements as a
        function of time (BJD\,--\,245\,0000.0). No correlation with
        radial velocity and time is observed suggesting that the
        Doppler signal is induced by the planet. }
\label{W56_Vspan}%
\end{figure}

   \begin{figure}
     \centering
     \includegraphics[width=0.5\textwidth]{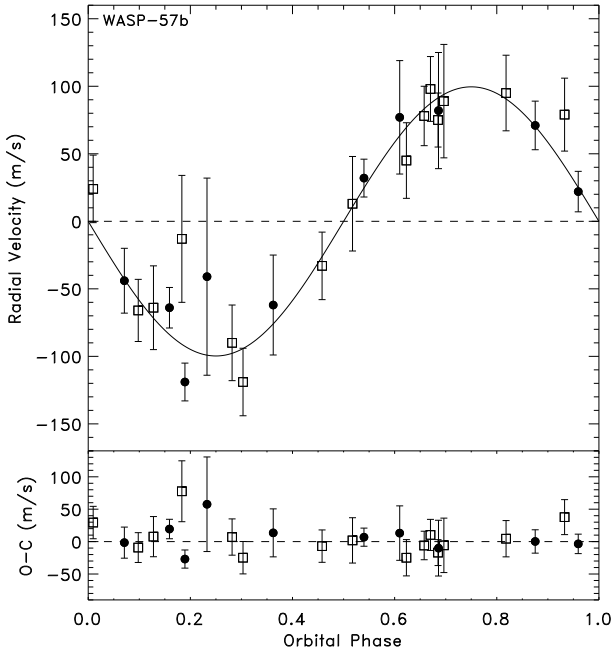}
     \caption{{\em Upper panel}: Similar to Figures \ref{W54_RVplot}
       and \ref{W56_RVplot} for WASP-57. The centre-of-mass velocity
       for each data set was subtracted from the RVs (\gammaS$ =
       -23.214$~\kms\ and \gammaC $= -23.228$~\kms). {\em Lower
         panel}: Residuals from the RV orbital fit plotted against
       time.} \label{W57_RVplot}%
\end{figure}

   \begin{figure}
     \centering
     \includegraphics[width=0.5\textwidth]{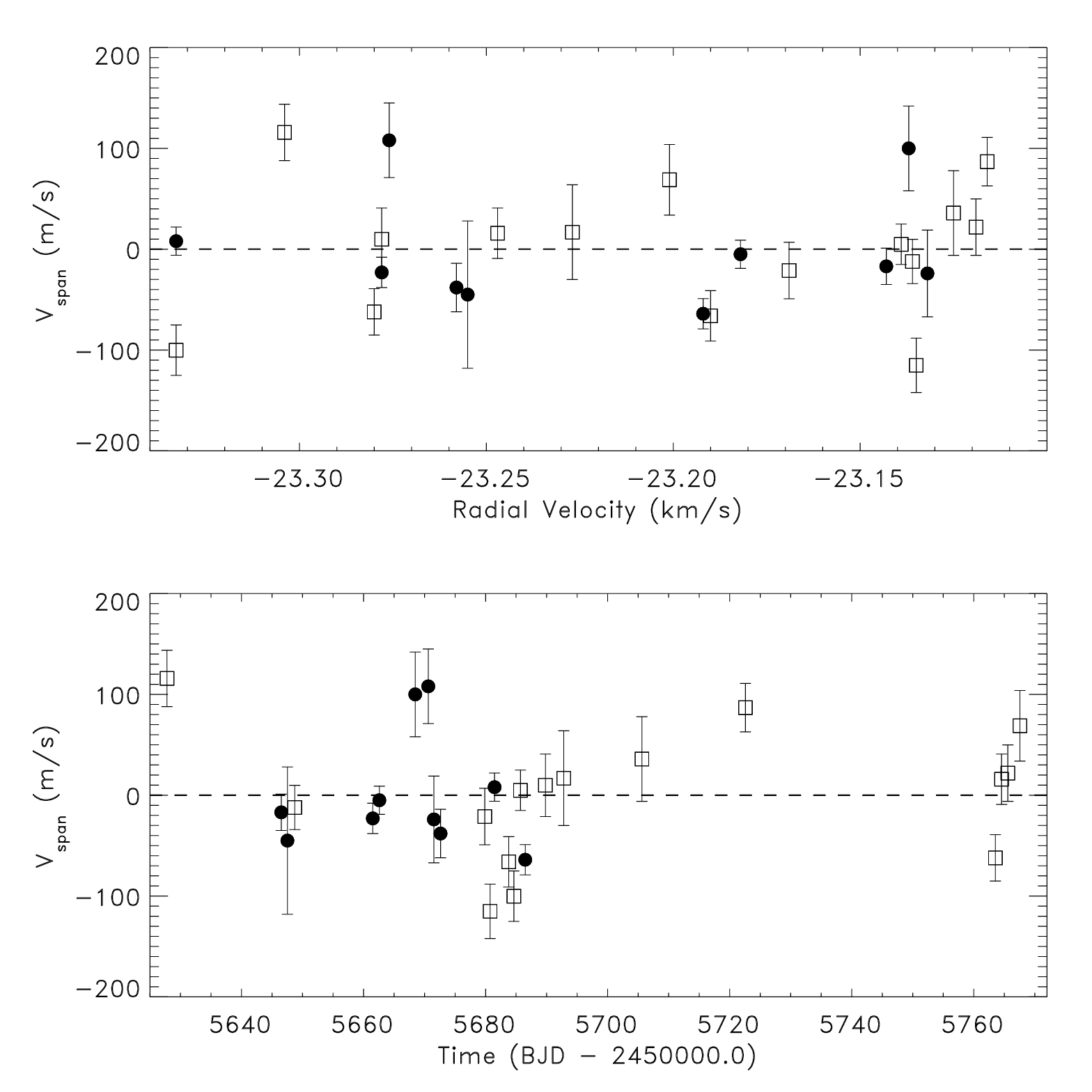}
   \caption{{\em Upper panel}: Same as Figures \ref{W54_Vspan} and
     \ref{W56_Vspan} we show the bisector span measurements for
     WASP-57 as a function of radial velocity, values are shifted to a
     zero-mean($<$$V_{span}$$>_{SOPHIE} = -2$~m\,s$^{-1}$,
     $<$$V_{span}$$>_{CORALIE} = 22$~m\,s$^{-1}$). {\em Lower panel}:
     The bisector span measurements as a function of time (BJD\,--\,
     245~0000.0). No correlation with radial velocity and time is
     observed suggesting that the Doppler signal is induced by the
     planet.} \label{W57_Vspan}%
 \end{figure}
 For all Figures presented in the paper we adopted the convention for
 which SOPHIE data are always represented as filled circles and
 CORALIE data are represented as open squares. In Figures
 \ref{W54_RVplot} to \ref{W57_Vspan} we present the RVs, ${\rm
   V}_{span}$, and the residuals O\,--\,C diagrams for the three
 systems. Both CORALIE and SOPHIE data sets are offset with respect to
 the radial velocity zero point, \gammaS~and \gammaC~, respectively
 (see Tables \ref{W54_params} and \ref{W56_params}). We examined
 V$_{span}$ to search for asymmetries in spectral line profiles that
 could result from unresolved binarity or indeed stellar
 activity. Such effects would cause the bisector spans to vary in
 phase with radial velocity. For the three systems no significant
 correlation is observed between the radial velocity and the line
 bisector, or the bisector and the time at which observation were
 taken. This supports each signal's origin as being planetary, rather
 than due to a blended eclipsing
 binary system, or to stellar activity (see \citealt{Queloz2001}). \\

\noindent
- {\bf WASP-54}'s follow up spectroscopy was obtained from both the
SOPHIE and CORALIE spectrographs (see Figures~\ref{W54_RVplot} and
\ref{W54_Vspan}). The $RMS$ for SOPHIE and CORALIE radial velocity
residuals to the best-fit model are $RMS_{\rm SOPHIE} =
33.6$~m\,s$^{-1}$ and $RMS_{\rm CORALIE}= 8.2$~m\,s$^{-1}$. Typical
internal errors for CORALIE and SOPHIE are of 10--15~m\,s$^{-1}$. The
significantly higher $RMS$ of the SOPHIE residuals is mostly due to
one observation (RV = 134~m\,s$^{-1}$).  Removing this measurement
results in a $RMS_{\rm SOPHIE}$ = 18~m\,s$^{-1}$, which is comparable
to the quoted internal error. We investigated the reasons of the
particularly large error bar associated with the measurement above
(55~m\,s$^{-1}$) and we found that it is due to a shorter exposure
time, cloud absorption, and Moon pollution. The specific observation
was obtained during grey time at a Moon distance of 57\degree.  To
estimate and remove the sky contamination we used the method described
in \citet{Pollacco2008} and \citet{Hebrard2008}, however, the RV shift
induced by the Moon was high (310~m\,s$^{-1}$) and the relative low
S/N resulted in a less accurate measurement (1.3-$\sigma$ away from
the residuals).

\noindent
- {\bf WASP-56} has radial velocity data only from SOPHIE (see
Figures~\ref{W56_RVplot} and \ref{W56_Vspan}).  The $RMS$ of the RV
residuals to the best-fit model is 19.4~m\,s$^{-1}$. When removing the
only discrepant RV value at phase 0.5 (RV = $-61$~m\,s$^{-1}$) the
overall $RMS$ reduces to 12~m\,s$^{-1}$, comparable to SOPHIE internal
error. 
\noindent
- Finally, for {\bf WASP-57} the $RMS$ of the SOPHIE and CORALIE
radial velocity residuals to the best-fit model are $RMS_{\rm SOPHIE}
= 22.3$~m\,s$^{-1}$ and $RMS_{\rm CORALIE}= 26.5$~m\,s$^{-1}$,
respectively (see Figures \ref{W57_RVplot} and \ref{W57_Vspan}). These
become 14~m\,s$^{-1}$ and 17.4~m\,s$^{-1}$ respectively for SOPHIE and
CORALIE data sets when ignoring the two measurements with the largest
errors.

\begin{figure}
   \centering
   \includegraphics[width=0.5\textwidth]{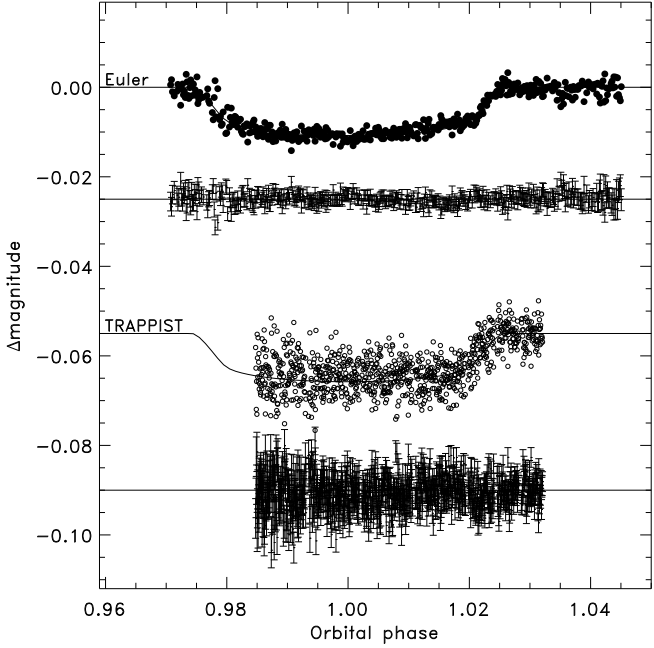}
   \caption{Euler $r-$band and TRAPPIST $`I+z'-$band follow up high
     signal-to-noise photometry of WASP-54 during the transit (see
     Table~\ref{tablephot}). The TRAPPIST light curve has been offset
     from zero by an arbitrary amount for clarity.  The data are
     phase-folded on the ephemeris from
     Table~\ref{W54_params}. Superimposed (black-solid line) is our
     best-fit transit model estimated using the formalism from
     \citet{Mandel2002}. Residuals from the fit are displayed
     underneath.}
     \label{W54_photFU}
    \end{figure}

\begin{figure}
   \centering
   \includegraphics[width=0.5\textwidth]{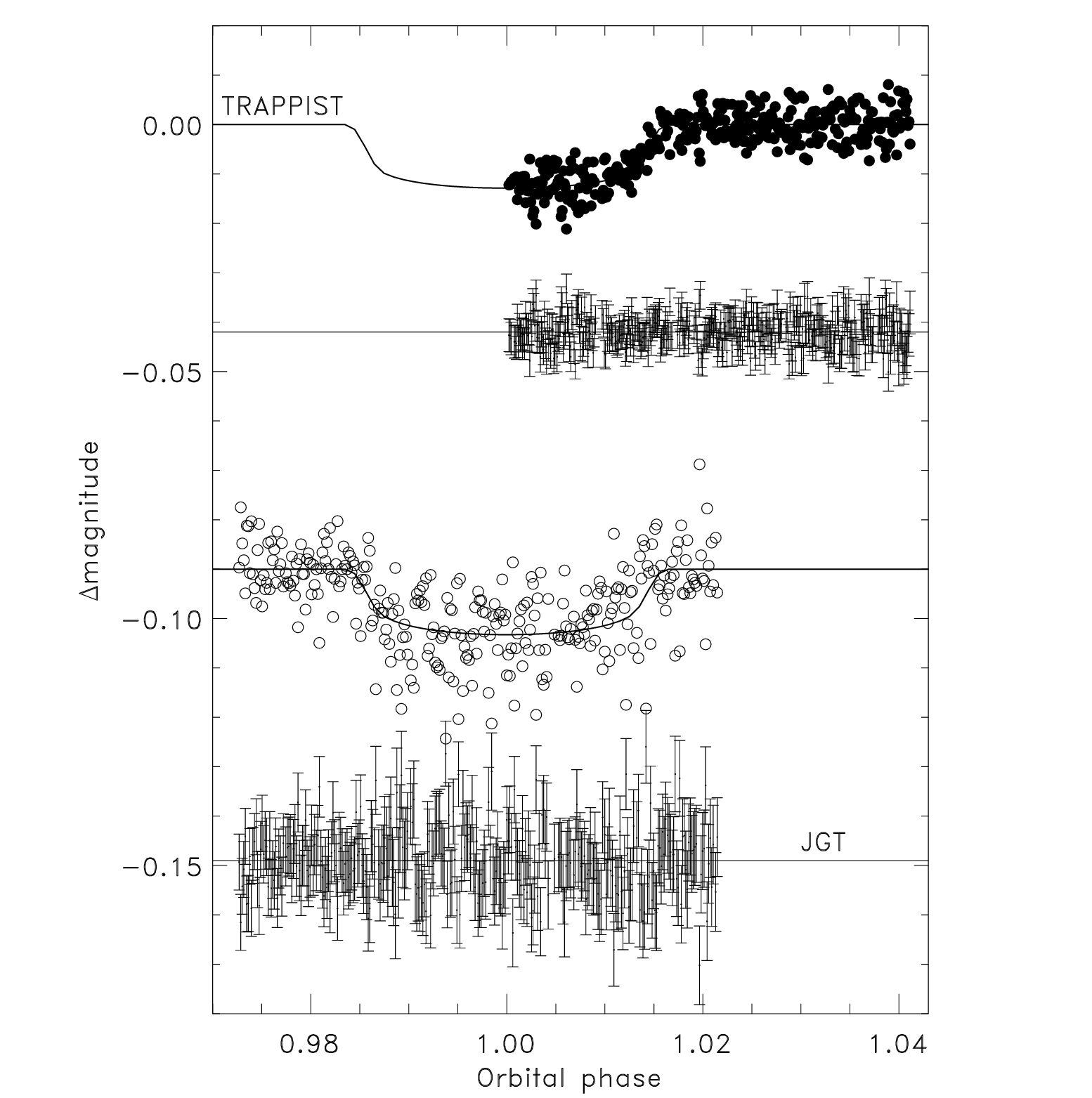}
   \caption{TRAPPIST $`I+z'-$band and JGT R-band follow up high
     signal-to-noise photometry of WASP-56 during the transit (see
     Table~\ref{tablephot}). The JGT light curve has been offset from
     zero by an arbitrary amount for clarity. The The data are
     phase-folded on the ephemeris from
     Table~\ref{W56_params}. Superimposed (black-solid line) is our
     best-fit transit model estimated using the formalism from
     \citet{Mandel2002}. Residuals from the fit are displayed
     underneath.}
     \label{W56_photFU}
    \end{figure}

\begin{figure}
   \centering
   \includegraphics[width=0.5\textwidth]{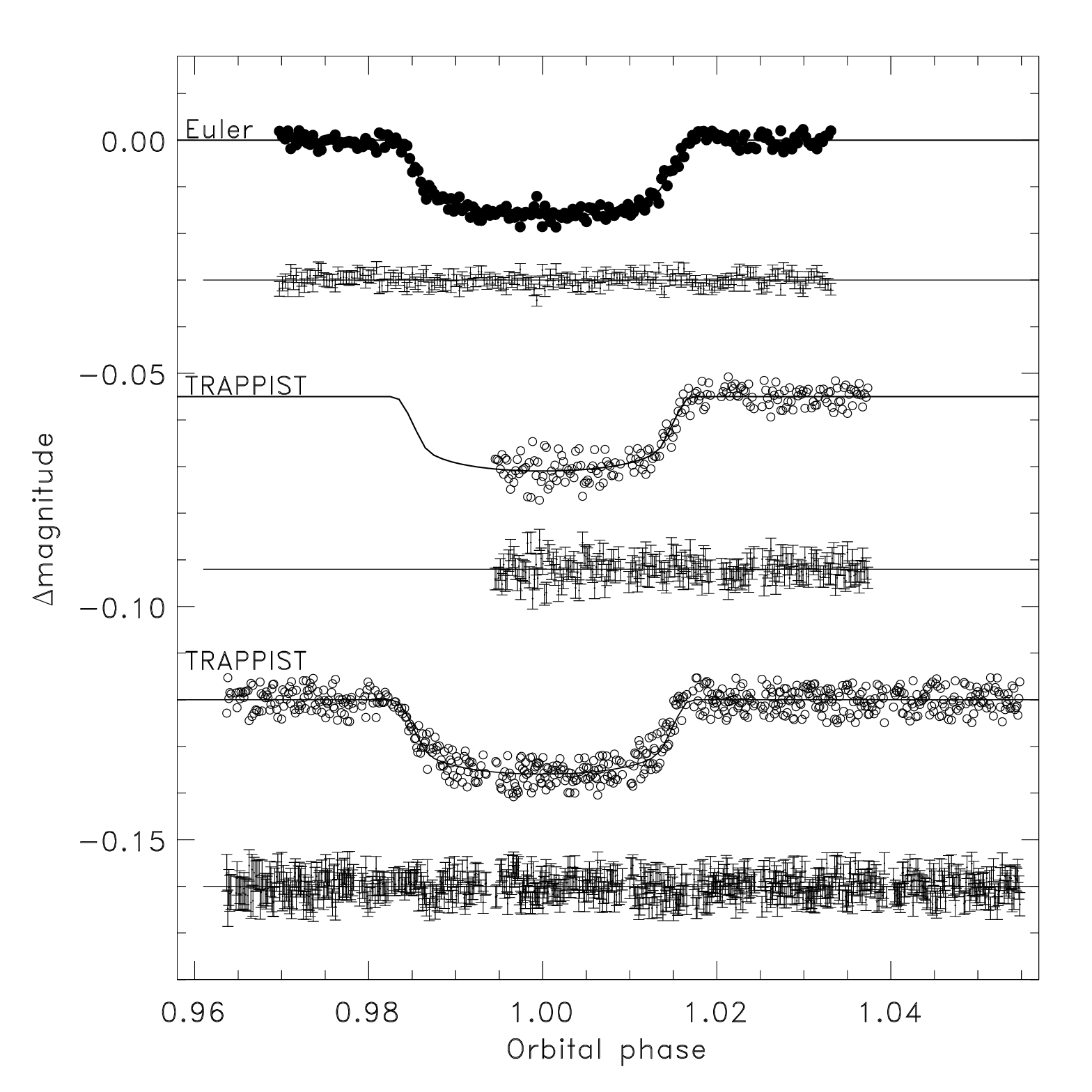}
   \caption{Euler $r-$band and TRAPPIST $`I+z'-$band follow up high
     signal-to-noise photometry of WASP-57 during the transit (see
     Table~\ref{tablephot}). The TRAPPIST light curves have been
     offset from zero by an arbitrary amount for clarity.  The data
     are phase-folded on the ephemeris from
     Table~\ref{W56_params}. Superimposed (black-solid lines) are the
     best-fit transit models estimated using the formalism from
     \citet{Mandel2002}. The residuals from each fit are displayed
     underneath the relative light curves.}
     \label{W57_photFU}
    \end{figure}

\subsection{follow up Multi-band Photometry}
In order to allow more accurate light curve modelling of the three new
WASP planets and tightly constrain their parameters, in-transit
high-precision photometry was obtained with the TRAPPIST and Euler
telescopes located at ESO La Silla Observatory in Chile. The TRAPPIST
telescope and its characteristics are described in \citet{Jehin2011}
and \citet{Gillon2011}. A detailed description of the physical
characteristics and instrumental details of EulerCam can be
found in \citet{Lendl2012}.

\begin{table}
\caption{Photometry for WASP-54, WASP-56 and WASP-57} 
%\begin{tabular*}{0.5\textwidth}{@{\extracolsep{\fill}}lccrl} 
\begin{tabular}{@{\extracolsep{\fill}}lccrl} 
%\begin{tabular}{|c|c|c|c|c|} 
\toprule
Planet & Date & Instrument & Filter& Comment \\
\midrule
{\multirow{2}{*}{WASP-54b}}& 06/04/2011 & EulerCam & Gunn $r$ 	& full transit \\
                           & 27/02/2012 & TRAPPIST & $I+z$ 	& partial transit\\
\midrule
{\multirow{2}{*}{WASP-56b}}& 16/05/2011 & TRAPPIST & $I+z$ 	& partial transit \\ 	
			   & 11/03/2012 & JGT	   & $R$	& full transit\\
\midrule
{\multirow{3}{*}{WASP-57b}}& 05/05/2011 & TRAPPIST & $I+z$ 	& partial transit \\
			   & 10/06/2011 & TRAPPIST & $I+z$ 	& full transit \\
			   & 10/06/2011 & EulerCam & Gunn $r$ 	& full transit\\
\bottomrule
\end{tabular}
\label{tablephot}
\end{table} 

All photometric data presented here are available from the NStED
database \footnote{http://nsted.ipac.caltech.edu}. One full and one partial
transits of WASP-54b have been observed by EulerCam in 2011 April 6 and
TRAPPIST in 2012 February 26, respectively. Only a partial transit of
WASP-56b was observed by TRAPPIST in 2011 May 16, and a full transit was
observed by JGT in 2012 March 11. A partial and a full transit of
WASP-57b were captured by TRAPPIST on the nights of 2011 May 5 and
June 10 respectively, while a full transit of WASP-57b was observed
with EulerCam in 2011 June 10. A summary of these observations is given
in Table~\ref{tablephot}.

We show in Figures~\ref{W54_photFU}, \ref{W56_photFU}, and \ref{W57_photFU}
the high S/N follow up photometry (EulerCam and TRAPPIST) for WASP-54b,
WASP-56b and WASP-57b respectively. In each plot we show the
differential magnitude versus orbital phase, along with the residual
to the best-fit model. The data are phase folded on the ephemerides 
derived by our analysis of each individual object (see
\S~\ref{results}). In Figures \ref{W54_photFU} and \ref{W57_photFU}
some of the light curves are assigned an arbitrary magnitude offset for
clarity.

\subsection{TRAPPIST  `I + z'--band photometry}
TRAPPIST photometry was obtained using a readout mode of
$2\times2$~MHz with $1\times1$ binning, resulting in a readout time of
6.1~s and readout noise 13.5 e$^-$pix$^{-1}$, respectively. A slight
defocus was applied to the telescope to optimise the observation
efficiency and to minimise pixel to pixel effects. TRAPPIST uses a
special `I+z' filter that has a transmittance $>90\%$ from 750~nm to
beyond 1100~nm. The positions of the stars on the chip were maintained
to within a few pixels thanks to the `software guiding' system that
regularly derives an astrometric solution to the most recently
acquired image and sends pointing corrections to the mount, if needed
(see e.g., \citealt{Gillon2011} for more details). A standard
pre-reduction (bias, dark, flat field correction), was carried out and
the stellar fluxes were extracted from the images using the
IRAF/DAOPHOT \footnote{IRAF is distributed by the National Optical
  Astronomy Observatory, which is operated by the Association of
  Universities for Research in Astronomy, Inc., under cooperative
  agreement with the National Science Foundation.}  aperture
photometry software \citet{Stetson1987}. After a careful selection of
reference stars differential photometry was then obtained.

\subsection{Euler r--band photometry}
Observations with the Euler-Swiss telescope were obtained in the Gunn
$r$ filter. The Euler telescope employs an absolute tracking system
which keeps the star on the same pixel during the observation, by
matching the point sources in each image with a catalogue, and
adjusting the telescope pointing between exposures to compensate for
drifts \citep{Lendl2012}. WASP-54b's observations were carried out
with a 0.2~mm defocus and one-port readout with exposure time of
30~s. All images were corrected for bias and flat field effects and
transit light curve were obtained by performing relative aperture
photometry of the target and optimal bright reference stars. For
WASP-57b no defocus was applied, and observations were performed with
four-port readout, and 60~s exposures. Six reference stars were used
to perform relative aperture photometry to obtain the final light
curve.

\section{Results}

\subsection{Stellar parameters}\label{sec:specsynth}
For all the three systems the same stellar spectral analysis has been
performed, co-adding individual CORALIE and SOPHIE spectra with a
typical final S/N of $\sim$80:1. The standard pipeline reduction
products were used in the analysis, and the analysis was performed
using the methods given in \citet{Gillon2009}. The \halpha\ line was
used to determine the effective temperature (\teff). The surface
gravity (\logg) was determined from the Ca~{\sc i} lines at 6122{\AA},
6162{\AA} and 6439{\AA} along with the Na~{\sc i} D and Mg~{\sc i} b
lines.  The elemental abundances were determined from equivalent width
measurements of several clean and unblended lines. A value for
micro-turbulence (\mictrb) was determined from Fe~{\sc i} using the
method of \cite{Magain1984}. The quoted error estimates include that
given by the uncertainties in \teff, \logg\, and \mictrb, as well as
the scatter due to measurement and atomic data uncertainties. The
projected stellar rotation velocity (\vsini) was determined by fitting
the profiles of several unblended Fe~{\sc i} lines. For each system a
value for macro-turbulence (\mactrb) was assumed based on the
tabulation by \cite{Bruntt2010b}, and we used the telluric lines
around 6300\AA~to determine the instrumental FWHM. The values for the
\mactrb~and the instrumental FWHM are given in
Table~\ref{Stellar_params}. There are no emission peaks evident in the
Ca H+K lines in the spectra of the three planet host stars. For each
stellar host the parameters obtained from the analysis are listed in
Table~\ref{Stellar_params} and discussed below:\\

{\bf WASP-54:} Our spectral analysis yields the following results:
$\teff=6100 \pm 100$~K, $\logg=4.2 \pm 0.1$ (cgs), and \feh $=-0.27
\pm 0.08$ dex, from which we estimate a spectral type F9. WASP-54's
stellar mass and radius were estimated using the calibration of
\citet{Torres2010}. We find no significant detection of lithium in the
spectrum of WASP-54, with an equivalent width upper limit of 0.4~m\AA,
corresponding to an abundance upper limit of $\log A$(Li) $<$ 0.4
$\pm$ 0.08. The non-detection of lithium together with the low
rotation rate obtained from \vsini ($P_{\rm rot} = 17.60 \pm 4.38$~d),
assuming $i^{\star}$ is perpendicular to the line of sight (thus
\vsini=V$_{equatorial}$), and the lack of stellar activity (shown by
the absence of Ca~{\sc ii} H and K emission), all indicate that the
star is relatively old. From the estimated \vsini\ we derived the
stellar rotation rates, and we used the expected spin-down timescale
(\citealt{Barnes2007}) to obtain a value of the stellar age through
gyrochronology. We estimate an age of $4.4^{+7.4}_{-2.7}$~Gyr, This
value also suggest the system is old. Although we point out that in
the case of WASP-54 using gyrochoronlogy to constrain the age of the
system could be inappropriate as the planet could have affected the
stellar rotation velocity via tidal interaction (see section \S 4.1
for more details). However, we note that the gyrochronological age we
obtain is in agreement with that from theoretical evolutionary models
discussed below, which imply that WASP-54 has evolved off the main sequence.\\

{\bf WASP-56 and WASP-57:} Both stellar hosts are of spectral type
G6V. From our spectral analysis we obtain the following parameters:
$\teff=5600 \pm 100$~K, and $\logg=4.45 \pm 0.1$ (cgs) for WASP-56,
$\teff=5600\pm100$~K, and $\logg=4.2\pm0.1$ (cgs) for WASP-57. As
before the stellar masses and radii are estimated using the
\citet{Torres2010} calibration. With a metallicity of \feh $=0.12$~dex
WASP-56 is more metal rich than the sun, while our spectral synthesis
results for WASP-57 show that it is a metal poor star (\feh
$=-0.25$~dex). For both stars the quoted lithium abundances take account
non-local thermodynamic equilibrium corrections
\citep{Carlsson1994}. The values for the lithium abundances if these
corrections are neglected are as follows: $\log A$(Li) = 1.32 and
$\log A$(Li) = 1.82 for WASP-56 and WASP-57, respectively. These
values imply an age of $\ga$ 5 Gyr for the former and an age of $\ga$
2 Gyr for the latter \citep{Sestito2005}. From \vsini\ we derived the
stellar rotation period $P_{\rm rot} = 32.58 \pm 18.51$~d for WASP-56,
implying a gyrochronological age \citep{Barnes2007} for the system of
$\sim 5.5^{+10.6}_{-4.6}$~Gyr. Unfortunately, the gyrochronological
age can only provide a weak constraint on the age of WASP-56. For
WASP-57 we obtain a rotation period of $P_{\rm rot} = 18.20 \pm
6.40$~d corresponding to an age of $\sim 1.9^{+2.4}_{-1.2}$~Gyr. Both
the above results are in agreement with the stellar ages obtained from
theoretical evolution models (see below) and suggest that WASP-56 is
quite old, while WASP-57 is a relatively young system.

\begin{table}
\caption{Stellar parameters of WASP-54, WASP-56, and WASP-57 from spectroscopic analysis.}
\begin{tabular}{@{\extracolsep{\fill}}lccc} 
%\begin{tabular}{lccc} 
  \toprule \hline
  Parameter	&WASP-54 	&WASP-56	&WASP-57 \\ 
  \hline
  \teff~(K)	&$6100\pm100$	&$5600\pm100$	&$5600\pm100$\\       
  \logg		&$4.2\pm0.1$ 	&$4.45\pm0.1$	&$4.2\pm0.1$ \\          
  \mictrb~(\kms)	&$1.4\pm0.2$	&$0.9\pm0.1$ 	&$0.7\pm0.2$\\     
  \vsini~(\kms)	&$4.0\pm0.8$	&$1.5\pm0.9$ 	&$3.7\pm1.3$\\     
  {[Fe/H]}	&$-0.27\pm0.08$ &$0.12\pm0.06$	&$-0.25\pm0.10$ \\     
  {[Na/H]}	&$-0.30\pm0.04$ &$0.32\pm0.14$	&$-0.20\pm0.07$ \\     
  {[Mg/H]}	&$-0.21\pm0.05$ &$0.24\pm0.06$	&$-0.19\pm0.07$ \\     
  {[Si/H]}	&$-0.16\pm0.05$ &$0.31\pm0.07$	&$-0.13\pm0.08$ \\     
  {[Ca/H]}	&$-0.15\pm0.12$ &$0.09\pm0.12$	&$-0.21\pm0.11$ \\     
  {[Sc/H]}	&$-0.06\pm0.05$ &$0.35\pm0.13$	&$-0.08\pm0.05$ \\     
  {[Ti/H]}	&$-0.16\pm0.12$ &$0.18\pm0.06$	&$-0.18\pm0.07$ \\     
  {[Cr/H]}	&$-0.21\pm0.12$ &$0.20\pm0.11$	&--   \\                     
  {[Co/H]}	& --        	&$0.35\pm0.10$	&--     \\                   
  {[Ni/H]}	&$-0.29\pm0.08$ &$0.21\pm0.07$	&$-0.25\pm0.10$ \\     
  $\log A$(Li)	&$< 0.4\pm0.08$ &$1.37\pm0.10$	&$1.87\pm0.10$ \\        
  Mass~(\msun)	&$1.15\pm0.09$	&$1.03\pm0.07$ 	&$1.01\pm0.08$ \\  
  Radius~(\rsun)	&$1.40\pm0.19$	&$0.99\pm0.13$	&$1.32\pm0.18$  \\  
  Sp. Type   	&F9		&G6		&G6 \\                     
  Distance~(pc)   &$200\pm30$ 	&$255\pm40$ 	&$455\pm80$ \\ 
  \hline
  \bottomrule
  \\
\end{tabular}
\label{Stellar_params}
\newline {\bf Note:} Mass and radius estimate using the
\cite{Torres2010} calibration. Spectral type estimated from \teff\
using the table in \cite{Gray2008}.
\end{table}

\begin{figure}
   \centering
   \includegraphics[angle=-90, width=0.49\textwidth]{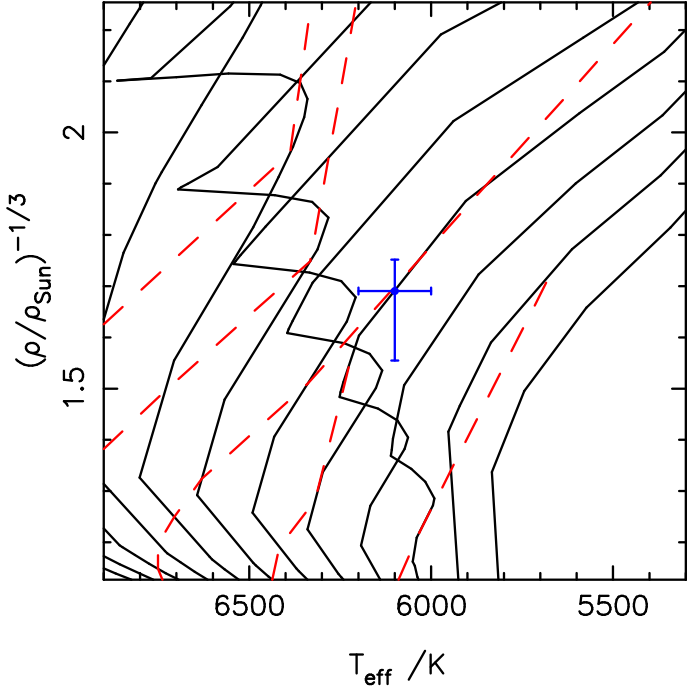}
   \caption{Isochrone tracks from \citet{Marigo2008} and
     \citet{Girardi2010} for WASP-54 using the metallicity \feh
     $=-0.27$ dex from our spectral analysis and the best-fit stellar
     density 0.2~\rhosun.  From left to right the solid lines are for
     isochrones of: 1.0, 1.3, 1.6, 2.0, 2.5, 3.2, 4.0, 5.0, 6.3, 7.9,
     10.0 and 12.6~Gyr. From left to right, dashed lines are for mass
     tracks of: 1.4, 1.3, 1.2, 1.1 and 1.0~\msun.}
    \label{iso1}%
   \end{figure}

\begin{figure}
   \centering
   \includegraphics[angle=-90, width=0.49\textwidth]{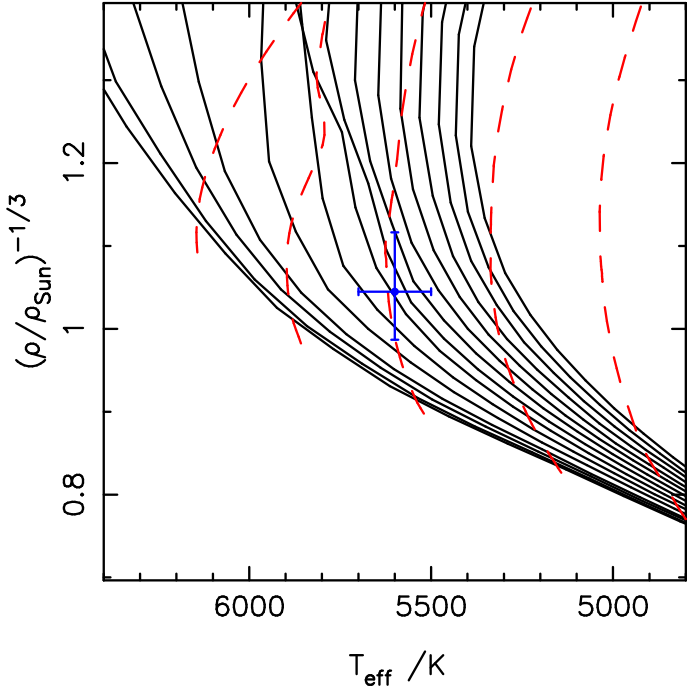}
   \caption{Isochrone tracks from \citet{Demarque2004} for WASP-56
     using the metallicity \feh $=0.12$ dex from our spectral analysis
     and the best-fit stellar density 0.88~\rhosun.  From left to
     right the solid lines are for isochrones of: 1.8, 2.0, 2.5, 3.0,
     4.0, 5.0, 6.0, 7.0, 8.0, 9.0, 10.0, 11.0, 12.0, 13.0 and
     14.0~Gyr. From left to right, dashed lines are for mass tracks
     of: 1.2, 1.1, 1.0, 0.9 and 0.8~\msun.}
    \label{iso2}%
   \end{figure}

\begin{figure}
   \centering
   \includegraphics[angle=-90, width=0.49\textwidth]{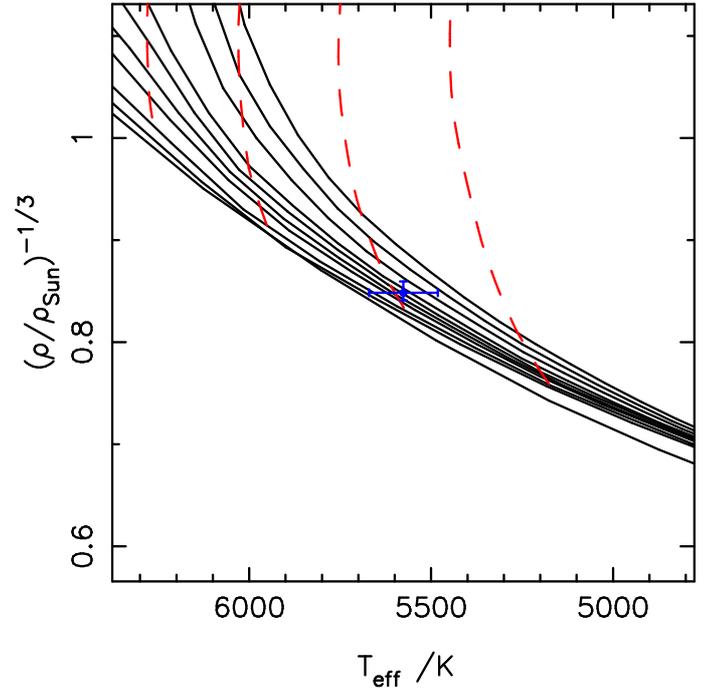}
   \caption{Isochrone tracks from \citet{Demarque2004} for WASP-57
     using the metallicity \feh $=-0.25$ dex from our spectral
     analysis and the best-fit stellar density 1.638~\rhosun.  From
     left to right the solid lines are for isochrones of: 0.1, 0.6,
     1.0, 1.6, 2.0, 2.5, 3.0, 4.0, 5.0 and 6.0~Gyr. From left to
     right, dashed lines are for mass tracks of: 1.1, 1.0, 0.9 and
     0.8~\msun.}
    \label{iso3}%
   \end{figure}

   For each system we used the stellar densities \rhostar, measured
   directly from our Markov-Chain Monte Carlo (MCMC) analysis (see \S
   \ref{results}, and also \citealt{Seager2003}), together with the
   stellar temperatures and metallicity values derived from
   spectroscopy, in an interpolation of four different stellar
   evolutionary models. The stellar density,~\rhostar, is directly
   determined from transit light curves and as such is independent of
   the effective temperature determined from the spectrum
   \citep{Hebb2009}, as well as of theoretical stellar models (if
   $\mpl \ll \mstar$ is assumed). Four theoretical models were used:
   a) the Padova stellar models (\citealt{Marigo2008}, and
   \citealt{Girardi2010}), b) the Yonsei-Yale (YY) models
   \citep{Demarque2004}, c) the Teramo models \citep{Pietrinferni2004}
   and finally d) the Victoria-Regina stellar models (VRSS)
   \citep{VandenBerg2006}. In Figures \ref{iso1}, \ref{iso2}, and
   \ref{iso3}, we plot the inverse cube root of the stellar density
   \rhostar$^{-1/3} =$~\rstar/\mstar$^{1/3}$ (solar units) against
   effective temperature, \teff, for the selected model mass tracks
   and isochrones, and for the three planet host stars
   respectively. For WASP-54 and WASP-56 the stellar properties
   derived from the four sets of stellar evolution models
   (Table~\ref{tracks}) agree with each other and with those derived
   from the \citet{Torres2010} calibration, within their 1--$\sigma$
   uncertainties. For WASP-57 the best-fit \mstar~from our MCMC
   analysis agrees with the values derived from theoretical stellar
   tracks with the exception of the Teramo models. The latter give a
   lower stellar mass value of $0.87\pm0.04$~\msun\ which is more than
   1--$\sigma$ away from our best-fit result (although within
   2--$\sigma$). The stellar masses of planet host stars are usually
   derived by comparing measurable stellar properties to theoretical
   evolutionary models, or from empirical calibrations. Of the latter,
   the most widely used is the \citet{Torres2010} calibration, which
   is derived from eclipsing binary stars, and relates \logg\ and
   \teff\ to the stellar mass and radius.  However, while \teff\ can
   be determined with high precision, \logg\ is usually poorly
   constrained, and thus stellar masses derived from the spectroscopic
   \logg\ can have large uncertainties and can suffer from
   systematics. For example the masses of 1000 single stars, derived
   by \citet{Valenti1998} via spectral analysis, were found to be
   systematically 10\% larger than those derived from theoretical
   isochrones.  A similar discrepancy was also found in the analysis
   of the stellar parameters of WASP-37 \citep{Simpson2011}, WASP-39
   \citep{Faedi2011}, and WASP-21 \citep{Bouchy2010}. Additionally,
   different sets of theoretical models might not perfectly agree with
   each other \citep{Southworth2010}, and moreover at younger ages
   isochrones are closely packed and a small change in \teff\ or
   \rhostar\ can have a significant effect on the derived stellar
   age. For each planet host star we show a plot with one set of
   stellar tracks and isochrones, while we give a comprehensive list
   of the four models' results in Table~\ref{tracks}.
   Using the metallicity of \feh$ =-0.27$ dex our best-fit stellar
   properties from the Padova isochrones (\citealt{Marigo2008} and
   \citealt{Girardi2010}) for WASP-54 yield a mass of
   $1.1^{+0.1}_{1.0}$\msun\ and a stellar age of
   6.3$^{+1.6}_{-2.4}$~Gyr, in agreement with the gyrochronological
   age and a more accurate estimate. The Padova isochrones together
   with the stellar mass tracks and WASP-54 results are shown In
   Figure \ref{iso1}. According to the stellar models, a late-F star
   with \feh$ =-0.27$ dex, of this radius and mass has evolved off the
   zero-age main sequence and is in the shell hydrogen burning phase
   of evolution with an age of 6.3$^{+1.6}_{-2.4}$~Gyr. The best-fit
   stellar ages from the other sets of stellar models of WASP-54 also
   agree with our conclusion. In Figure \ref{iso1} the large
   uncertainty on the minimum stellar mass estimated from
   interpolation of the Padova isochrones is likely due to the
   proximity to the end of the main sequence kink. The Padova
   evolutionary models were selected nevertheless, because they show
   clearly the evolved status of WASP-54.

   In Figures \ref{iso2}, and \ref{iso3} we show the best-fit
   Yonsei-Yale stellar evolution models and mass tracks
   \citep{Demarque2004} for the planet host stars WASP-56 and WASP-57,
   respectively. Using the metallicity of \feh$ = 0.12$ dex for
   WASP-56 our fit of the YY-isochrones gives a stellar mass of
   $1.01^{+0.03}_{-0.04}$ \msun\ and a stellar age of
   $6.2^{+3.0}_{-2.1}$~Gyr. This is in agreement with the Li abundance
   measured in the spectral synthesis (see
   Table~\ref{Stellar_params}), and supports the conclusion that
   WASP-56 is indeed an old system. Using the metallicity of \feh$ =
   -0.25$ dex derived from our spectral analysis of WASP-57, we
   interpolate the YY-models and we obtain a best-fit stellar mass of
   $0.89^{+0.04}_{-0.03}$~\msun and age of
   $2.6^{+2.2}_{-1.8}$~Gyr. These results also agree with our results
   from spectral synthesis and shows that WASP-57 is a relatively
   young system. For each system the uncertainties in the derived
   stellar densities, temperatures and metallicities were included in
   the error calculations for the stellar ages and masses, however
   systematic errors due to differences between various evolutionary
   models were not considered.

\subsection{Planetary parameters}\label{results}
The planetary properties were determined using a simultaneous
Markov-Chain Monte Carlo analysis including the WASP photometry, the
follow up TRAPPIST and Euler photometry, together with SOPHIE and
CORALIE radial velocity measurements (as appropriate see
Table~\ref{tablephot} and Tables~\ref{W54_RVtable},
\ref{W56_RVtable}, and \ref{W57_RVtable}). A detailed description of
the method is given in \citet{Cameron2007} and \citet{Pollacco2008}.
Our iterative fitting method uses the following parameters: the epoch
of mid transit $T_{0}$, the orbital period $P$, the fractional change
of flux proportional to the ratio of stellar to planet surface areas
$\Delta F = R_{\rm pl}^2/R_{\star}^2$, the transit duration $T_{14}$,
the impact parameter $b$, the radial velocity semi-amplitude $K_{\rm
  1}$, the stellar effective temperature \teff\ and metallicity \feh,
the Lagrangian elements \secos\ and \sesin~(where $e$ is the
eccentricity and $\omega$ the longitude of periastron), and the
systematic offset velocity $\gamma$. For WASP-54 and WASP-57 we fitted
the two systematic velocities $\gammaC$ and $\gammaS$ to allow for
instrumental offsets between the two data sets. The sum of the
$\chi^2$ for all input data curves with respect to the models was used
as the goodness-of-fit statistic. For each planetary system four
different sets of solutions were considered: with and without the
main-sequence mass-radius constraint in the case of circular orbits
and orbits with floating eccentricity.

An initial MCMC solution with a linear trend in the systemic velocity
as a free parameter, was explored for the three planetary systems,
however no significant variation was found. For the treatment of the
stellar limb-darkening, the models of \citet{Claret2000, Claret2004}
were used in the $r$-band, for both WASP and Euler photometry, and in
the $z$-band for TRAPPIST photometry.

From the parameters mentioned above, we calculate the mass $M$, radius
$R$, density $\rho$, and surface gravity $\log g$ of the star (which
we denote with subscript $_{\star}$) and the planet (which we denote
with subscript $_{\rm pl}$), as well as the equilibrium temperature of
the planet assuming it to be a black-body ($T_{\rm
  pl,A=0}$) and that energy is efficiently redistributed from the
planet's day-side to its night-side. We also calculate the transit
ingress/egress times $T_{\rm 12}$/$T_{\rm 34}$, and the orbital
semi-major axis $a$. These calculated values and their 1--$\sigma$
uncertainties from our MCMC analysis are presented in
Tables~\ref{W54_params} and \ref{W56_params} for WASP-54, WASP-56 and
WASP-57. The corresponding best-fitting transit light curves are shown
in Figures~\ref{W54_waspLC}, \ref{W56_waspLC}, and \ref{W57_waspLC} and in
Figures~\ref{W54_photFU}, \ref{W56_photFU}, and \ref{W57_photFU}. The
best-fitting RV curves are presented in Figures~\ref{W54_RVplot},
\ref{W56_RVplot}, and \ref{W57_RVplot}.\\

-- For {\bf WASP-54} the MCMC solution imposing the main sequence
mass-radius constraint gives unrealistic values for the best-fit
stellar temperature and metallicity, as we expected for an evolved
star. We then relaxed the main sequence constraint and explored two
solutions: one for a circular and one for an eccentric orbit. In the
case of a non-circular orbit we obtain a best-fit value for $e$ of
$0.067^{+0.033}_{-0.025}$. This is less than a 3--$\sigma$ detection,
and as suggested by \citet{LucySweeney1971}, Eq. 22, it could be
spurious. From our analysis we obtain a best-fit $\chi^2$ statistic of
$\chi^{2}_{\rm circ} = 24.3$ for a circular orbit, and $\chi^{2}_{\rm
  ecc} = 18.6$ for an eccentric orbit. The circular model is
parameterised by three parameters: K, \gammaS\, and \gammaC, while the
eccentric model additionally constrains \ecos\ and \esin. We used the
23 RV measurements available and we performed the Lucy \& Sweeney
F-test (Eq. 27 of \citealt{LucySweeney1971}), to investigate the
probability of a truly eccentric orbit for WASP-54b. We obtained a
probability of 9\% that the improvement in the fit produced by the
best-fitting eccentricity could have arisen by chance if the orbit
were real circular. \citet{LucySweeney1971} suggest a 5\% probability
threshold for the eccentricity to be significant. From our MCMC
analysis we obtain a best-fit value for $\omega =
62^{+20}_{-30}$~degree, this differs from 90~\degree\ or 270~\degree\
values expected from an eccentric fit of a truly circular orbit (see
\citealt{Laughlin2005}).  We decided to investigate further our
chances to detect a truly eccentric orbit which we discuss in \S
\ref{W54_ecc}. Table~\ref{W54_params} shows our best-fit MCMC
solutions for WASP-54b for a forced circular orbit, and for an orbit
with floating eccentricity. However, based on our analysis in \S
\ref{W54_ecc}, we adopted the eccentric solution.\\

-- For {\bf WASP-56b} the available follow up spectroscopic and
photometric data do not offer convincing evidence for an eccentric
orbit. The free-eccentricity MCMC solution yields a value of $e =
0.098\pm0.048$. The \citet{LucySweeney1971} F-test, indicates that
there is a 42\% probability that the improvement in the fit could have
arisen by chance if the orbit were truly circular. With only a partial
high S/N follow up light curve it is more difficult to precisely
constrain the stellar and planetary parameters (e.g., the time of
ingress/egress, the impact parameter $b$, and $a$/\rstar), however the
full, although noisy, low S/N GJT light curve (see Figure
\ref{W56_photFU}), allows us to better constrain the parameters
mentioned above. Therefore we decided to relax the main-sequence
constrain on the stellar mass and radius and we adopt a circular orbit.\\

-- For {\bf WASP-57b} the follow up photometry and radial velocity
data allowed us to relax the main-sequence mass-radius constrain and
perform an MCMC analysis leaving the eccentricity as free parameter.
However, our results do not show evidence for an eccentric orbit, and
the Lucy \& Sweeney test yields a 100\% probability that the orbit is
circular. Moreover, we find that imposing the main-sequence constraint
has little effect on the MCMC global solution. Thus, we decided to
adopt no main sequence prior and a circular orbit.

%%%%%%%%%%%%%%%%%%%%%%%
%MCMC parameters 
%%%%%%%%%%%%%%%%%%%%%%%
\begin{table} 
\caption{System parameters of WASP-54} 
\label{W54_params} 
%\begin{tabular*}{0.5\textwidth}{@{\extracolsep{\fill}}lc} 
\begin{tabular}{lcc} 
\hline 
Parameter (Unit) & Value&Value  \\
\\
& Circular Solution & Eccentric Solution \\ 
\hline 
\\
$P$ (d) 		 & $3.693649^{+0.000013}_{-0.000009}$          & $3.6936411^{+0.0000043}_{-0.0000074}$\\
$T_{0}$ (BJD) 		 & $2455522.04373^{+0.00079}_{-0.00071}$       & $2455518.35087^{+0.00049}_{-0.00056}$\\
$T_{\rm 14}$ (d) 	 & $0.1882^{+0.0023}_{-0.0031}$                & $0.1863^{+0.0015}_{-0.0018}$ \\
$T_{\rm 12} = T_{\rm 34}$ (d) \medskip 	& $0.0221^{+0.0019}_{-0.0038}$ & $0.0203^{+0.0009}_{-0.0011}$  \\
$\Delta F=\rpl^{2} / \rstar^{2}$ \medskip& $0.0088\pm0.0003$           & $0.0086\pm0.0002$ \\
$b$ \medskip 		 & $0.537^{+ 0.044}_{- 0.131}$                 & $0.490^{+0.026}_{-0.044}$ \\
$i$ ($^\circ$) \medskip  & $84.8^{+ 1.6}_{- 0.6}$                      & $84.97^{+0.63}_{-0.59}$ \\
$K_{\rm 1}$ (m\,s$^{-1}$)& $73\pm 2$                                   & $73\pm2$\\
$\gammaC$  (\kms) 	 & $-3.1335 \pm 0.0004$                        & $-3.1345 \pm0.0009 $\\
$\gammaS$ (\kms) 	 & $-3.1109 \pm 0.0004$                        & $-3.1119\pm0.0009$\\
$e\cos\omega$            & 0 (fixed)                                   & $0.030^{+0.021}_{-0.022}$\\
$e\sin\omega$            & 0 (fixed)                                   & $0.055^{+0.037}_{-0.036}$\\
$e$ 		 	 & 0 (fixed)                                   & $0.067^{+0.033}_{-0.025}$\\
$\omega$ ($^\circ$)      & 0 (fixed)                                   & $62^{+21}_{-33}$\\
$\phi_{\rm mid-occultation}$& 0.5                                      & $0.519^{+0.013}_{-0.014}$\\
$T_{\rm 58}$ (d)            & ---                                      & $0.20 \pm 0.01$\\
$T_{\rm 56}=T_{\rm 78}$ (d) \medskip & ---                             &$0.0232^{+0.0032}_{-0.0023}$\\
\mstar~(\msun) 		 & $1.201^{+0.034}_{-0.036}$                   &$1.213 \pm 0.032$\\
\rstar~(\rsun) 		 & $1.80^{+0.07}_{-0.16}$                      &$1.828^{+0.091}_{-0.081}$\\
$\log g_{\star}$ (cgs) \medskip & $4.01^{+0.07}_{-0.03}$               &$3.997^{+0.032}_{-0.035}$\\
\rhostar~(\rhosun) \medskip 	& $0.21^{+ 0.06}_{- 0.02}$             &$0.198^{+0.025}_{-0.024}$\\
\mpl~(\mj) 			& $0.626 \pm 0.023$                    &$0.636^{+0.025}_{-0.024}$\\
\rpl~(\rj) \medskip		& $1.65^{+0.09}_{-0.18}$               &$1.653^{+0.090}_{-0.083}$\\
$\log g_{\rm pl}$ (cgs) \medskip & $2.724^{+ 0.088}_{- 0.042}$         &$2.726\pm0.042$\\
\rhopl~(\rhoj) 			 & $0.14^{+0.05}_{-0.02}$              &$0.141^{+0.022}_{-0.019}$\\
$a$ (AU)  \medskip		 & $0.0497 \pm 0.0005$                 &$0.04987 \pm0.00044$ \\
$T_{\rm pl, A=0}$ (K) 		 & $1742^{+ 49}_{- 69}$                &$1759 \pm 46$\\
\\ 
\hline 
\multicolumn{2}{l}{$^{a}$ $T_{\rm 14}$: time between 1$^{st}$ and 4$^{th}$ contact} 
\end{tabular} 
\tablefoot{\rj/\rsun = 0.10273; \mj/\msun=0.000955}
\end{table}

%%%%%%%%%%%%%%%%%%%%%%%
%MCMC paramters 
%%%%%%%%%%%%%%%%%%%%%%%
\begin{table} 
\caption{System parameters of WASP-56 and WASP-57} 
\label{W56_params} 
%\begin{tabular*}{0.5\textwidth}{@{\extracolsep{\fill}}lc} 
\begin{tabular}{lcc} 
\hline 
&WASP-56&WASP-57\\
\\%\cmidrule(c){2-3}
Parameter (Unit) & Value&Value  \\ 
\hline 
\\
$P$ (d) 				& $4.617101^{+0.000004}_{-0.000002}$	& $2.838971\pm0.000002$\\        
$T_{0}$ (BJD) 				& $2455730.799\pm0.001$			& $2455717.87811\pm0.0002$\\    
$T_{\rm 14}$ (d) 			& $0.1484\pm0.0025$			& $0.0960\pm0.0005$\\     
$T_{\rm 12}=T_{\rm 34}$ (d) \medskip 	& $0.0146\pm0.0005$			& $0.01091^{+0.00032}_{-0.00018}$\\  
$\Delta F=\rpl^{2}/\rstar^{2}$ \medskip & $0.01019\pm0.00041$			& $0.01269\pm0.00014$\\         
$b$ \medskip 				& $0.272^{+0.029}_{-0.018}$		& $0.345^{+0.033}_{-0.014}$\\   
$i$ ($^\circ$) \medskip 		& $88.5^{+0.1}_{-0.2}$			& $88.0^{+0.1}_{-0.2}$\\      
$K_{\rm 1}$ (m\,s$^{-1}$) 		& $69\pm 4$				& $100\pm 7$\\       
\gammaS~(\kms) 			        & $4.6816 \pm 0.0001$			& $-23.214 \pm 0.002$\\ 
\gammaC~(\kms)) 			& --- 					& $-23.228 \pm 0.002$\\
$e$ 					& 0 (fixed)				& 0 (fixed)\\               
\mstar~(\msun) 				& $1.017 \pm 0.024$			& $0.954\pm0.027$\\ 
\rstar~(\rsun) 				& $1.112^{+0.026}_{-0.022}$		& $0.836^{+0.07}_{-0.16}$\\          
$\log g_{\star}$ (cgs) \medskip		& $4.35\pm0.02$				& $4.574^{+0.009}_{-0.012}$\\ 
\rhostar~(\rhosun) \medskip 		& $0.74\pm0.04$ 			& $1.638^{+0.044}_{-0.063}$\\ 
\mpl~(\mj) 				& $0.571^{+0.034}_{-0.035}$		& $0.672^{+0.049}_{-0.046}$\\  
\rpl~(\rj) \medskip			& $1.092^{+0.035}_{-0.033}$		& $0.916^{+0.017}_{-0.014}$\\     
$\log g_{\rm pl}$ (cgs) \medskip        & $3.039^{+0.035}_{-0.038}$		& $3.262^{+0.063}_{-0.033}$\\  
\rhopl~(\rhoj) 				& $0.438^{+0.048}_{-0.046}$		& $0.873^{+0.076}_{-0.071}$\\
$a$ (AU)  \medskip			& $0.05458 \pm 0.00041$ 		& $0.0386\pm 0.0004$\\
$T_{\rm pl, A=0}$ (K) 		        & $1216^{+25}_{-24}$			& $1251^{+21}_{-22}$\\  
\\   
\hline 
\multicolumn{2}{l}{$^{a}$ $T_{\rm 14}$: time between 1$^{st}$ and 4$^{th}$ contact} 
\end{tabular} 
\tablefoot{\rj/\rsun = 0.10273; \mj/\msun=0.000955}
\end{table}

\subsection{The Eccentricity of WASP-54b}\label{W54_ecc}
Here we investigate possible biases in the detection of the
eccentricity of WASP-54b, and we explore the possibility that the
eccentricity arises from the radial velocity measurements alone. It is
well known that eccentricity measurements for a planet in a circular
orbit can only overestimate the true zero eccentricity
\citep{Ford2006b}. We want to quantify whether using only the radial
velocity measurements at hand we can find a significant difference in
the best-fit model of truly circular orbit compared to that of a real
eccentric orbit with $e = 0.067$, as suggested by our free-floating
eccentric solution. Indeed, the best-fit eccentricity depends on the
signal-to-noise of the data, on gaps in the phase coverage, on the
number of orbital periods covered by the data set, and the number of
observations (see e.g., \citealt{Zakamska2011}).

We use the uncertainty of the CORALIE and SOPHIE radial velocity
measurements of WASP-54, the MCMC best-fit orbital period, velocity
semi-amplitude $K$, and epoch of the transit T$_{0}$ as initial
parameters, to compute synthetic stellar radial velocities at each
epoch of the actual WASP-54 RV data set. We generated synthetic radial
velocity data using the Keplerian model of \citet{Murray1999}, for the
two input eccentricities, $e=0$ and $e=0.067$. We then added Gaussian
noise deviates to the synthetic RV at each epoch, corresponding to the
original RV uncertainties added in quadrature with 3.5~m\,s$^{-1}$
accounting for stellar jitter. In this way at each observation time
the simulated velocity is a random variable normally distributed
around a value {\it v(t$_i$)} + $\gamma$, with dispersion
$\sqrt{\sigma^{2}_{\rm obs}+\sigma^{2}_{\rm Jitter}}$, where $\gamma$
is the centre of mass velocity. In this manner the simulated data have
similar properties to the real WASP-54 velocities but with the
advantage of having a known underlying eccentricity and orbital
properties.

\begin{figure}
  \centering
   \includegraphics[width=0.5\textwidth]{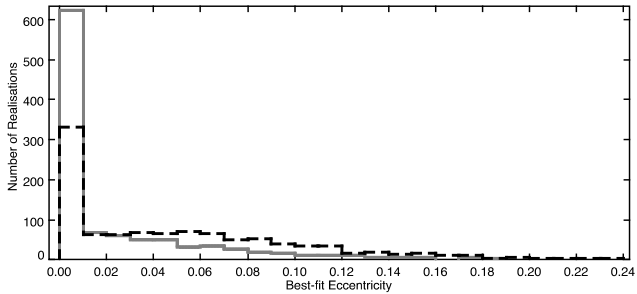}
   \caption{Histograms of the output eccentricity distributions for
     the input $e = 0$ (grey solid line), and for the input
     eccentricity of 0.067 (black dashed line). }
\label{distribution}
\end{figure}

\begin{figure}
   \centering
   \includegraphics[width=0.5\textwidth]{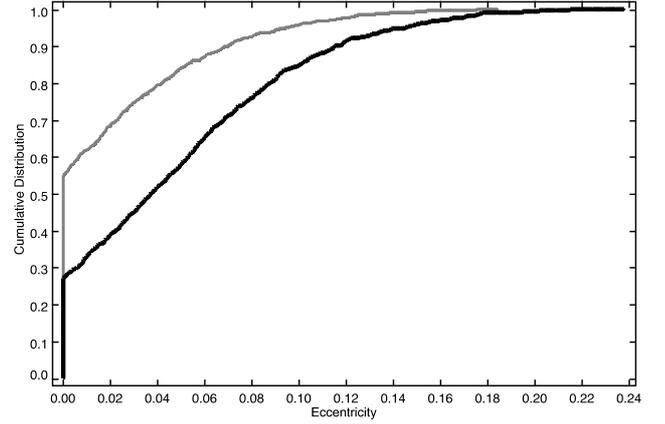}
   \caption{Cumulative Distribution Function for the two sets of
     best-fit eccentricities. We show in grey the CDFs for the
     simulated data sets with underline circular orbits, and in black
     the CDFs for the eccentric ones. }
\label{CDF}
\end{figure}

We generated 1000 synthetic data sets for each input eccentricity and
found the best-fit values for $e$.~In Figure \ref{distribution} we
show the output eccentricity distributions for the input $e = 0$ (grey
solid line), and for the input eccentricity of 0.067 (black dashed
line). Clearly, the one-dimensional distribution of the output
eccentricity is highly asymmetric. Because $e$ is always a positive
parameter, the best-fit eccentricities are always positive values. We
used the 1000 output best-fit values of $e$ of the two samples of
synthetic data sets to perform the Kolmogorov-Smirnov (KS) test to
asses our ability to distinguish between the two underlying
distributions. In Figure~\ref{CDF} we show the Cumulative Distribution
Function (CDFs) of the 1000 mock best-fit eccentricities for the two
cases. We show in grey the CDFs for the simulated data sets with
underlying circular orbits, and in black the CDFs for the eccentric
case. We calculated D, the absolute value of the maximum difference
between the CDFs of the two samples, and we used tabulated values for
the KS test. We are able to reject the hypothesis that the two samples
have the same underlying distribution with a confidence of 99.999\%.
We then conclude that the detected eccentricity of WASP-54b could
indeed be real. We point out however, that time-correlated noise
could potentially yield a spurious eccentricity detection. This is
difficult to asses with the limited number of radial velocity
observations at hand; more data and photometric monitoring during
transit and secondary eclipses are needed to better constrain the
orbital parameters of WASP-54b. In the following, we adopt the
eccentric MCMC model for WASP-54b.

\section{Discussion}
We report the discovery of three new transiting extra-solar planets
from the WASP survey, WASP-54b, WASP-56b and WASP-57b. In the
following we discuss the implications of these new planet discoveries.

\subsection{WASP-54b}
From our best-fit eccentric model we obtain a planetary mass of
$0.634^{+0.025}_{-0.024}$~\mj~ and a radius of
1.653$^{+0.090}_{-0.083}$~\rj~which yields a planetary density of
0.141$^{+0.022}_{-0.019}$~\rhoj. Thus, WASP-54b is among the least
dense, most heavily bloated exoplanets and shares similarities with
low-density planets such as WASP-17b \citep{Anderson2010a}, WASP-31b
\citep{Anderson2010b}, and WASP-12b \citep{Hebb2010}. These exoplanets
have short orbital periods, orbit F-type host stars and therefore are
highly irradiated. Using standard coreless models from
\citet{Fortney2007} and \citet{Baraffe2008}, we find that WASP-54b has
a radius more than 50\% larger than the maximum planetary radius
predicted for a slightly more massive 0.68\mj\ coreless planet,
orbiting at 0.045~AU from a 5~Gyr solar-type star
(R$_{expected}$=1.105~\rj\ ).  However, WASP-54 is an F-type star and
therefore hotter than the Sun, implying that WASP-54b is more strongly
irradiated. The low stellar metallicity ([Fe/H]$ = -0.27\pm0.08$) of
WASP-54 supports the expected low planetary core-mass thus favouring
radius inflation. Different mechanisms have been proposed to explain
the observed anomalously large planetary radii such as tidal heating
\citep{Bodenheimer2001,Bodenheimer2003}, kinetic heating
\citep{Guillot2002}, enhanced atmospheric opacity \citep{Burrows2007},
and semi-convection \citep{Chabrier2007}. While each individual
mechanism would presumably affect all hot Jupiters to some degree --
for example the detected non-zero eccentricity of WASP-54b and the
strong stellar irradiation are contributing to the radius inflation --
they cannot explain the entirety of the observed radii
(\citealt{Fortney2010}, \citealt{Baraffe2010}), and additional
mechanisms are needed to explain the inflated radius of WASP-54b. More
recently, \citet{Batygin2011} and \citet{Perna2010} showed that the
ohmic heating mechanism (dependent on the planet's magnetic field and
atmospheric heavy element content), could provide a universal
explanation of the currently measured radius anomalies (see also
\citealt{Laughlin2011}). However, according to \citet{Wu2012} Eq.6,
the maximum expected radius for WASP-54b, including ohmic heating, is
1.61\rj. This value for the radius is calculated assuming a system's
age of 1~Gyr and that ohmic heating has acted since the planet's
birth.~Therefore, if we regard this value as an upper limit for the
expected radius of WASP-54b at 6~Gyr, it appears more difficult to
reconcile the observed anomalously large radius of WASP-54b (although
the value is within 1-$\sigma$) even when ohmic heating is considered,
similarly to the case of WASP-17b \citep{Anderson2010b}, and HAT-P-32b
\citep{Hartman2011}, as discussed by \citet{Wu2012}.  Additionally,
\citet{Huang2012} find that the efficacy of ohmic heating is reduced
at high \teff\, and that it is difficult to explain the observed radii
of many hot Jupiters with ohmic heating under the influence of
magnetic drag. The ability of ohmic heating in inflating planetary
radii depends on how much power it can generate and at what depth,
with deeper heating able to have a stronger effect on the planet's
evolution (\citealt{Rauscher2012}; \citealt{Guillot2002}).
\citet{Huang2012} models predict a smaller radius for WASP-54b (see
their Fig. 12).

\begin{figure}
   \centering
   \includegraphics[width=0.5\textwidth]{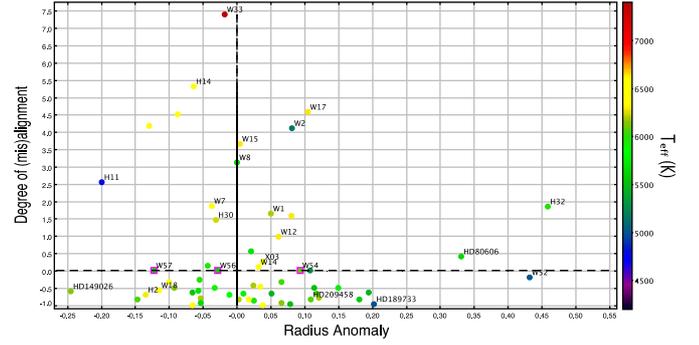}     
   \caption{ We plot the degree of (mis)alignment, $\eta_{\rm RM}$,
     versus the radius anomaly $\mathcal R$ and the stellar effective
     temperature for known planets (see text for details). The black
     dashed line separates aligned from mis-aligned systems following
     our description. The black solid line indicates $\mathcal
     R=0$. Known planetary systems with characteristics across the
     parameter space are indicated as follows: W for WASP, H for
     HatNet, C for CoRoT, K for Kepler, and others by their full
     name. Our three new discoveries are indicated by fuchsia square
     symbols. We note that no RM measurement is yet available for
     WASP-54b, WASP-56b, WASP-57b.}
\label{RM}
\end{figure}

However, the discrepancy between observations and the ohmic heating
models in particular in the planetary low-mass regime
(e.g. \citealt{Batygin2011}), shows that more understanding of
planets' internal structure, chemical composition and evolution is
required to remove assumptions limiting current theoretical
models. Moreover, \citet{Wu2012} suggest that ohmic heating can only
suspend the cooling contraction of hot--Jupiters; planets that have
contracted before becoming subject to strong irradiation, can not be
re-inflated. Following this scenario, the observed planetary radii
could be relics of their past dynamical histories. If this is true, we
could expect planets migrating via planet--planet scattering and/or
Kozai mechanisms (which can become important at later stages of
planetary formation compared to disc migration,
(\citealt{Fabrycky2007}, \citealt{Nagasawa2008}), to show a smaller
radius anomaly and large misalignments. This interesting possibility
can be tested by planets with Rossiter-McLaughlin (RM) measurements of
the spin--orbit alignment (\citealt{Holt1893}; \citealt{Rossiter1924};
\citealt{McLaughlin1924} \citealt{Winn2006}). We use all systems from
the RM-encyclopedia (http://ooo.aip.de/People/rheller/) to estimate
the degree of spin-orbit (mis)alignment. We consider aligned every
system with $|\lambda|<30\degree$ (a 3--$\sigma$ detection from zero
degrees; \citealt{Winn2010}), and define $\eta_{\rm RM} =
(|\lambda|-30\degree)/30\degree$ as the measure of the degree of
(mis)alignment of each system. This has the advantage to show all
aligned systems in the region $-1 < \eta_{\rm RM} \le 0$. In
Figure~\ref{RM} we show $\eta_{\rm RM}$ versus the radius anomaly and
the stellar temperature \teff\ (as a colour gradient) for planets with
RM measurements. We have calculated the radius anomaly, $\mathcal R$,
as follows (R$_{\rm obs} - {\rm R}_{\rm exp})/{\rm R}_{\rm exp}$, see
also \citet{Laughlin2011}. We find it difficult to identify any
correlation (see also \citealt{Jackson2012}, and their Fig. 11).  We
want to stress here that the uncertainty in the timescales of
planet--planet scattering and Kozai migration mechanism relative to
disc migration remain still large and thus any robust conclusion can
not be drawn until all the underlying physic of migration is
understood. Moreover, \citet{Albrecht2012} suggest that the Kozai
mechanism is responsible for the migration of the majority, if not
all, hot Jupiters, those mis-aligned as well as the aligned ones, and
finally, that tidal interaction plays a central role. Additionally,
measurements of spin--orbit obliquities could bear information about
the processes involved in star formation and disc evolution rather
than on the planet migration. For example \citet{Bate2010} have
recently proposed that stellar discs could become inclined as results
of dynamics in their environments (e.g. in stellar clusters), and
\citet{Lai2011} suggest that discs could be primordially mis-aligned
respect to the star, although \citet{Watson2011} could find no
evidence of disc mis-alignment.  Last but not least we note that not
many planets showing the radius anomaly have measured RM effects and
that more observations are needed to constrain theoretical models
before any robust conclusion can be drawn.

\begin{figure*}
   \centering
   \includegraphics[width=0.49\textwidth]{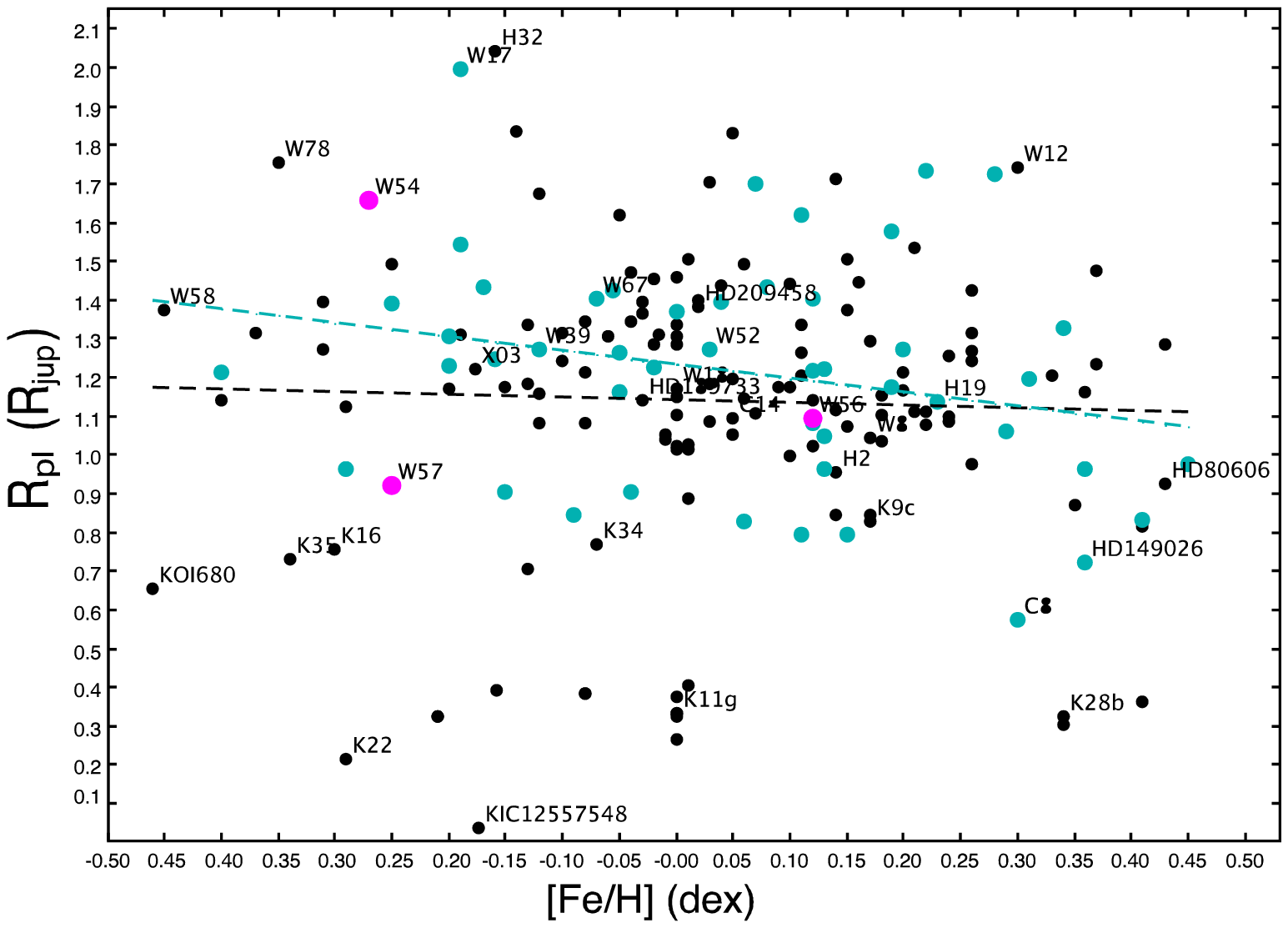}
   \includegraphics[width=0.49\textwidth]{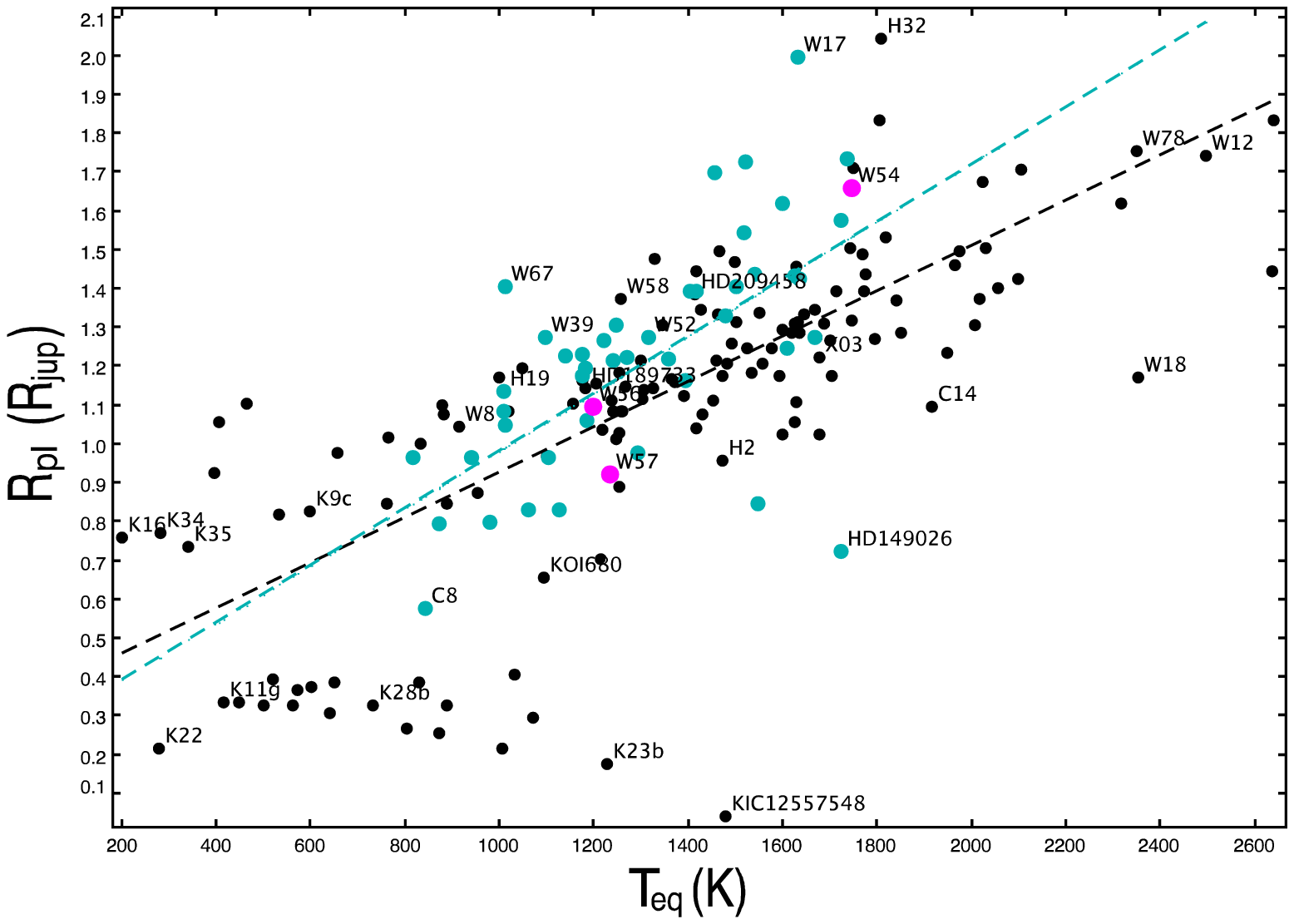}
   \caption{The Planetary radius versus stellar metallicity \feh\
     (dex)({\em left panel}), and as a function of planet equilibrium
     temperature T$_{eq}$ (K) ({\em right panel}). Black points
     indicate all planets, planets in the mass range
     0.2$<\mpl<$0.7\mj\, are indicated by Turquoise points.  Black and
     turquoise dashed lines are a simple linear regression to the two
     samples. Data were taken from the exoplanet encyclopaedia. Known
     planetary systems with characteristics across the parameter space
     are indicated as in Figure~\ref{RM}. Our three new discoveries
     are indicated by fuchsia filled circles.}
\label{Rpl_feh_Teq}
\end{figure*}

We investigate the radius anomaly of WASP-54b with respect to the full
sample of known exoplanets and respect to the sample of Saturn-mass
planets including the latest discoveries and the planets presented in
this work (for an updated list see the extra-solar planet
encyclopaedia \footnote{http://exoplanet.eu/}.  In Figure
\ref{Rpl_feh_Teq} we plot the planetary radius versus the stellar
metallicity ({\em left panel}), and as a function of the planet
equilibrium temperature T$_{\rm eq}$ ({\em right panel}). WASP-54b is
indicated with a filled fuchsia circle. We highlight the sample of
Saturn mass planets in turquoise; the black and turquoise dashed lines
show a simple linear regression for the full exoplanet and the Saturn
mass sample respectively.

\begin{figure*}
   \centering
   \includegraphics[width=0.49\textwidth]{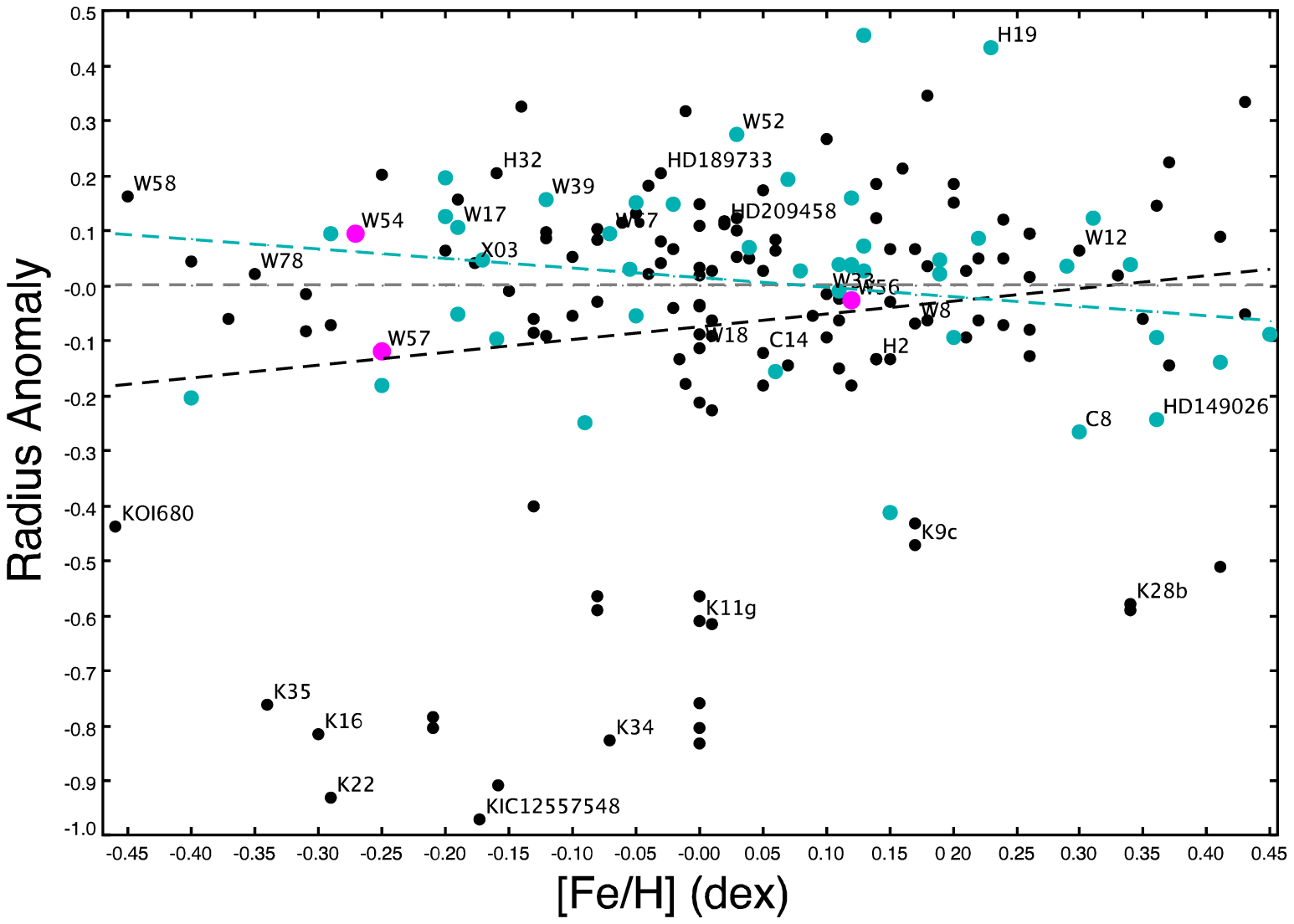}
   \includegraphics[width=0.49\textwidth]{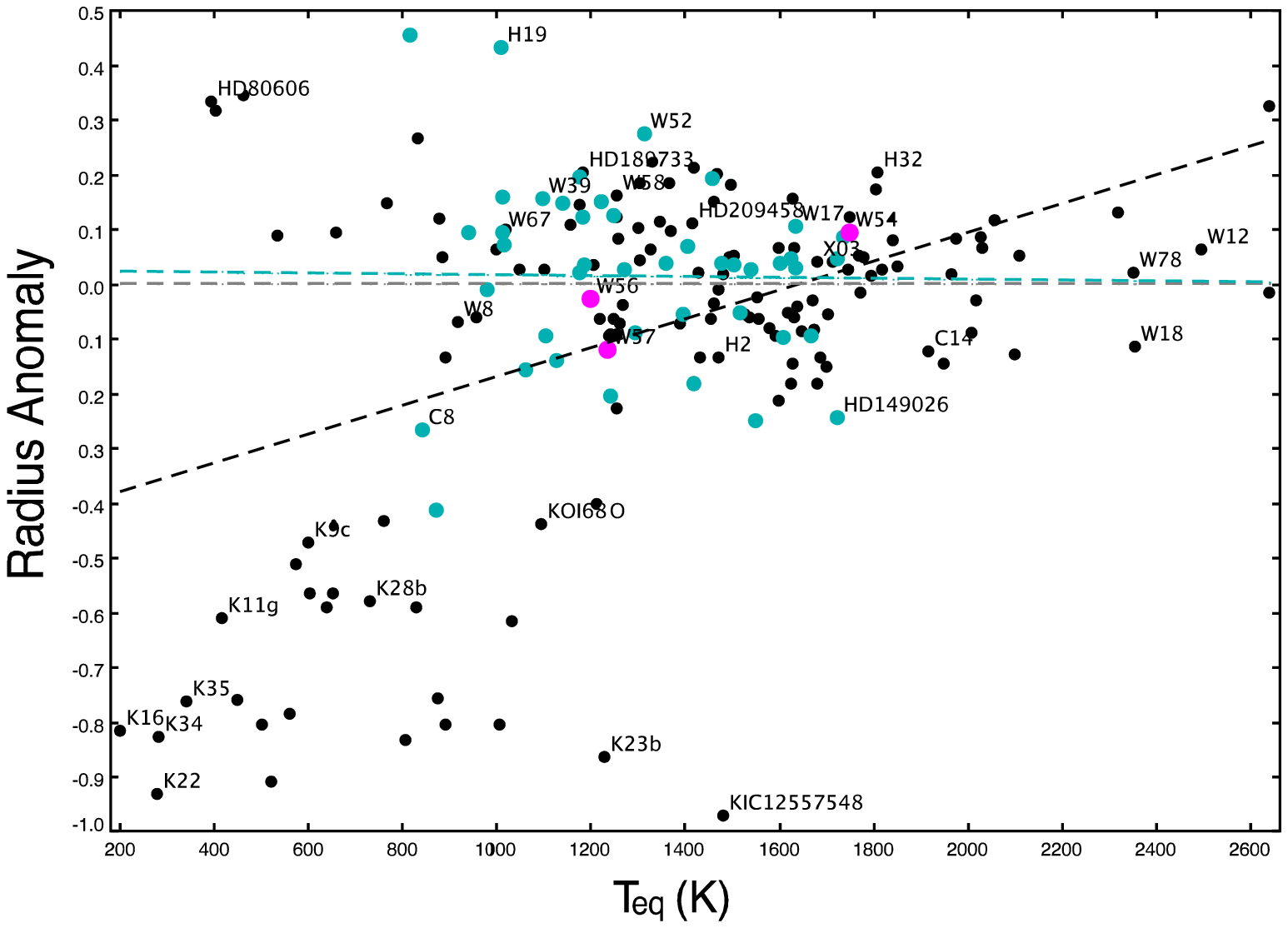}
   \caption{The radius anomaly, $\mathcal R$ versus \feh~(dec) ({\em
       left panel}), and as a function of T$_{\rm eq}$ (K) ({\em right
       panel}).  The radius anomaly is calculated using the
     \citet{Enoch2012} empirical relation for planetary radii. Symbols
     are like in Figure \ref{Rpl_feh_Teq}.  Our three new discoveries
     are indicated by fuchsia filled circles.}
\label{newRanomaly}
\end{figure*}

WASP-54b appears to strengthen the correlation between planets'
inflated radii and stellar temperature, and the anti-correlation with
metallicity. With an irradiation temperature of $\sim$2470~K, WASP-54b
is in the temperature region, identified by \citet{Perna2012}, with
T$_{\rm irr}>2000$~K (T$_{\rm irr}$ as defined by \citealt{Heng2012}),
in which planets are expected to show large day-night flux contrast
and possibly temperature inversion, in which case the ohmic power has
its maximum effect. With more gas giant planet detections we can start
to shed some light on which mechanism might be more efficient and in
which circumstances. For example, in the case of WASP-39b
\citep{Faedi2011}, and WASP-13b (\citealt{Skillen2009};
\citealt{Barros2011}), two Saturn-mass planets with similar density to
WASP-54b, but much less irradiated (T$_{\rm eq}=1116$~K, and T$_{\rm
  eq}=1417$~K, respectively) %; see Figure \ref{Ranomaly}),
ohmic heating could play a less significant role (see for example
\citealt{Perna2012}). However, many unknowns still remain in their
model (e.g., internal structure, magnetic field strength, atmospheric
composition). We selected all the exoplanets in the mass range between
$0.1<\mpl<12$\mj, and we used the empirical calibrations for planetary
radii derived by \citet{Enoch2012} to calculate the expected planetary
radius R$_{exp}$. We then used these values to derive again the Radius
Anomaly. We plot the results of radius anomaly versus stellar
metallicity \feh~and as a function of T$_{\rm eq}$ in Figure
\ref{newRanomaly}. Colours and symbols are like in Figure
\ref{Rpl_feh_Teq}, the dotted line indicates a zero radius anomaly.
The \citet{Enoch2012} relations take into account the dependence of
the planetary radius from planet T$_{\rm eq}$, \feh, and also tidal
heating and semi-major axis.  However, we note that even including
this dependence, there remained significant scatter in the observed
radii in particular for systems such as WASP-17b, WASP-21b and also
WASP-54b, WASP-56b and WASP-57b.

Thus, more gas giant planet discoveries and their accurate
characterisation are needed to compare planetary physical properties,
in order to understand their thermal structure and distinguish between
various theoretical models. With a magnitude of V$=10.42$ WASP-54 is a
bright target and thus its spectroscopic and photometric
characterisation is readily feasible. Given the detected non-zero
eccentricity we encourage secondary eclipse observations in the
IR. These observations will allow precise measurement of the system's
eccentricity as well as provide fundamental information on the thermal
structure of the planet. However, we note that even in the case of a
circular orbit our MCMC solution for WASP-54 yields a very similar,
inflated, planetary radius R$^{circ}_{pl}=1.65$ \rj, see Table
\ref{W54_params}.

Finally, WASP-54 is an old ($> 6$~Gyr) F9 star which has evolved off
the main sequence and is now in the Hydrogen shell-burning phase of
stellar evolution (see Figure~\ref{iso1}). This implies that recently
in its life WASP-54 has increased its radius by more than 60\%, and
thus it is ascending the red giant branch (RGB). WASP-54b thus could
be experiencing drag forces, both gravitational and tidal, which will
affect its orbital radius. Two main factors contribute to the change
of the planetary orbit: 1) the host-star can lose mass via stellar
wind which could be accreted by the planet resulting in an increase of
the orbital radius, and 2) the planet's orbital angular momentum
decreases due to the tidal drag, leading to a decrease of the orbital
radius. This is expressed by $\dot{a}_{\rm pl} = \dot{a}_{\rm tide} +
\dot{a}_{\rm mass\, loss}$ \citep{Zahn1977,Zahn1989}. While the second
term is negligible for Jupiter-like planets \citep{Duncan1998}, the
term due to tides can become important as it is proportional to
$(\rstar/a(r))^{8}$, where $a(r)$ is the decreasing orbital
radius. This could re-set the clock of WASP-54b making it appear
younger, and maybe contributing to the planet radius
inflation. Finally, any planet within the reach of the star's radius
during RGB and asymptotic giant branch (AGB) phases (about 1~AU for a
Solar-type star) will spiral-in and eventually merge with the star or
evaporate (\citealt{Villaver2007}; \citealt{Livio1984}).

\subsection{WASP-56b and WASP-57b}
Our modelling of the WASP-56 system yields a planet mass of
0.571$^{+0.034}_{-0.035}$~\mj\ and radius of
1.092$^{+0.035}_{-0.033}$~\rj\, which in turn give a planet density of
0.438$^{+0.048}_{-0.046}$~\rhoj. Hence WASP-56b belongs to the class
of Saturn-mass planets and does not show a radius anomaly
\citep{Laughlin2011}. Figure~\ref{Rpl_feh_Teq} shows that the radius
of WASP-56b is not inflated. With a metallicity of \feh = +0.12~dex
WASP-56 is more metal rich than the Sun, and with an age of
$\sim$6.3~Gyr standard planetary evolutionary models from
\citet{Fortney2007} and \citet{Baraffe2008}, show that WASP-56b has a
core of approximately $>10$~M$_{\oplus}$ of heavy material. Moreover,
WASP-56b orbits a main sequence G6 star, thus it is subject to less
stellar irradiation than WASP-54b. Our best-fit MCMC model for the
WASP-57 system yields a planetary mass of 0.672~\mj\ and a radius of
0.916~\rj. Similarly to WASP-56b, we find that WASP-57b has a high
density ($\rhopl=0.873~\rhoj$) and small radius.  As our analysis
suggests that WASP-57b may be relatively young ($\sim2.6$~Gyr), it may
posses a significant core-mass of more than 50~$M_{\oplus}$, as
derived from standard evolutionary models \citep{Fortney2007}.
Figure~\ref{Rpl_feh_Teq} shows that WASP-56 and WASP-57 systems have
different physical properties, for example the radius of WASP-57b
appears to depart from the observed trend with stellar metallicity,
and the planet possibly shares more similarities with the giant planet
in the HAT-P-12 system \citep{Hartman2009}. The derived radii of
WASP-56b and WASP-57b are also consistent with more recent planetary
models that include ohmic heating. Models from \citet{Batygin2011},
\citet{Wu2012}, and \citet{Huang2012}, all agree in that planets with
lower effective temperatures have smaller radii, and that Jupiter-mass
planets with \teff$<$1400~K experience no significant radius inflation
at all (\citealt{Miller2011}; and Figure 6
by\citealt{Batygin2011}). However, despite all available information,
we are still far from knowing the composition of the deep interior of
exoplanets. For example the very existence of planetary cores, their
masses, as well as the amount and distribution of heavy elements in
the planets' core or in their envelopes, remain
undetermined. Recently, \citet{Wilson2012} suggested that planetary
cores, mostly composed of rock and ices, can be eroded and/or dissolve
(depending on their mass) into the metallic H/He layers above, and
thus be redistributed in the planetary envelope (see also
\citealt{Umemoto2006}). This can have significant implications for
giant planets' thermal evolution, their
radius contraction, and overall structure. \\

In conclusion, it is clear that continued exoplanet discoveries are
needed to provide stronger constraints on theoretical models of close
in giant planets and hence their physical properties.

\begin{acknowledgements} 
  The SuperWASP Consortium consists of astronomers primarily from
  Queens University Belfast, St Andrews, Keele, Leicester, The Open
  University, Warwick, Isaac Newton Group La Palma and Instituto de
  Astrofsica de Canarias. The SuperWASP-N camera is hosted by the
  Issac Newton Group on La Palma and WASPSouth is hosted by SAAO. We
  are grateful for their support and assistance. Funding for WASP
  comes from consortium universities and from the UK's Science and
  Technology Facilities Council.  Spectroscopic observations were made
  with SOPHIE spectrograph mounted on the 1.9-m telescope at
  Observatoire de Haute-Provence (CNRS), France and at the ESO La
  Silla Observatory (Chile) with the CORALIE Echelle spectrograph
  mounted on the Swiss telescope.TRAPPIST is funded by the Belgian
  Fund for Scientific Research (Fonds National de la Recherche
  Scientifique, FNRS) under the grant FRFC 2.5.594.09.F, with the
  participation of the Swiss National Science Fundation
  (SNF). M. Gillon and E. Jehin are FNRS Research Associates. The
  research leading to these results has received funding from the
  European Community's Seventh Framework Programme (FP7/2007-2013)
  under grant agreement number RG226604 (OPTICON). FF is grateful to
  the anonymous referee for useful comments significantly improving
  the paper.
  
   \end{acknowledgements}

\bibliographystyle{aa}
\bibliography{Faedi}

\onecolumn

\begin{table} 
\caption{Radial velocity and line bisector span measurements of WASP-54.} 
%\begin{tabular*}{0.5\textwidth}{@{\extracolsep{\fill}}lccrl} 
\begin{tabular}{cccccrcr} 
\toprule
\toprule
BJD & RV & $\sigma_{\rm RV}$ & V$_{span}$& Instrument & ${\rm RV}-\gamma$ & V$_{span}-<$V$_{span}$$>$&O -- C\\
$-$2\,450\,000 &(\kms)&(\kms)&(\kms)&& (m\,s$^{-1}$)&(m\,s$^{-1}$)&(m\,s${^{-1}}$)\\
\midrule
5413.4899 &	$-$3.080 &	0.009 & 0.068    & CORALIE & 53		& $-$59	&	7\\
5596.8475 &	$-$3.202 &	0.007 & 0.032    & CORALIE & $-$69	&   1		&	4\\
5622.7168 &	$-$3.200 &	0.010 & 0.083    & CORALIE & $-$67	& $-$16	&	6\\
5623.7808 &	$-$3.109 &	0.027 & 0.070    & CORALIE & 25		&  13		&	5\\
5624.7570 &	$-$3.052 &	0.007 & 0.056    & CORALIE & 81		& $-$39	&	13\\
5626.7092 &	$-$3.191 &	0.008 & 0.031    & CORALIE & $-$58	& 113	&	4\\
5629.8659 &	$-$3.216 &	0.008 & 0.050    & CORALIE & $-$82	& $-$12	&	$-$15\\
5635.7802 &	$-$3.057 &	0.009 & 0.047    & CORALIE & 76		&  19		&	6\\
5637.8171 &	$-$3.187 &	0.010 & 0.062    & CORALIE & $-$53	& $-$17	&	7\\
5638.8200 &	$-$3.101 &	0.007 & 0.041    & CORALIE & 32		&  34		&	$-$16\\
5639.8881 &	$-$3.091 &	0.007 & 0.077    & CORALIE & 43		&  21		&	2\\
5646.7309 &	$-$3.061 &	0.007 & 0.027    & CORALIE & 72		&   7		&	0\\
5647.6789 &	$-$3.149 &	0.006 & 0.020    & CORALIE & $-$16	& $-$18	&	$-$9\\
5648.7057 &	$-$3.205 &	0.006 & 0.051    & CORALIE & $-$72	&   1		&	$-$2\\
5651.8347 &	$-$3.191 &	0.008 & 0.019    & CORALIE & $-$57	& $-$2	&	$-$1\\
5677.5887 &	$-$3.173 &	0.007 & 0.047    & CORALIE & $-$40	&  13		&	7\\
\hline
5646.4778 &	$-$3.067 &	0.012 &$-$0.030 & SOPHIE  & 67		& $-$8	&	0\\
5649.4440 &	$-$3.150 &	0.012 & 0.030	    & SOPHIE  & $-$17	&  28		&	$-$13\\
5659.5180 &	$-$3.199 &	0.012 & 0.013	    & SOPHIE  & $-$65	& $-$22	&	6\\
5663.5397 &	$-$3.179 &	0.016 & 0.042     & SOPHIE  & $-$46		& $-$29	&	22\\
5664.5255 &	$-$3.117 &	0.035 &$-$0.010 & SOPHIE  & 17		&   2		&	$-$16\\
5665.4957 &	$-$2.999 &	0.055 & 0.140     & SOPHIE  & 134		& $-$30	&	72\\
5668.4228 &	$-$3.110 &	0.025 & 0.017     & SOPHIE  & 23		& $-$2	&	$-$30\\
\bottomrule
\end{tabular}
\label{W54_RVtable}
\end{table}

\begin{table} 
\caption{Radial velocity and line bisector span measurements of WASP-56.} 
%\begin{tabular*}{0.5\textwidth}{@{\extracolsep{\fill}}lccrl} 
\begin{tabular}{cccccrcr} 
\toprule
\hline
BJD & RV & $\sigma_{\rm RV}$ & V$_{span}$& Instrument & ${\rm RV}-\gamma$ & V$_{span}-<$V$_{span}$$>$&O -- C\\
$-$2\,450\,000 &(\kms)&(\kms)&(\kms)&& (m\,s$^{-1}$)&(m\,s$^{-1}$)&(m\,s${^{-1}}$)\\
\midrule
5647.4063 & 4.724 & 0.012 & $-$0.045 & SOPHIE &	42	&$-$6	&16\\
5649.4028 & 4.626 & 0.010 & $-$0.045 & SOPHIE &	$-$56	&$-$5	&$-$5\\
5651.4189 & 4.721 & 0.035 & $-$0.057 & SOPHIE &	39 	&$-$17	&$-$22\\
5659.5384 & 4.723 & 0.011 & $-$0.039 & SOPHIE &	41 	&1	&14\\
5660.5028 & 4.745 & 0.010 & $-$0.028 & SOPHIE &	63 	&11	&$-$5\\
5668.4425 & 4.621 & 0.027 & $-$0.061 & SOPHIE &	$-$61 	&$-$22	&$-$58\\
5670.3468 & 4.716 & 0.011 & $-$0.020 & SOPHIE &	34 	&20	&$-$4\\
5671.5029 & 4.637 & 0.011 & $-$0.049 & SOPHIE &	$-$45 	&$-$9	&13\\
5672.4088 & 4.627 & 0.010 & $-$0.039 & SOPHIE &	$-$55 	&1	&1\\
5681.4453 & 4.605 & 0.011 & $-$0.038 & SOPHIE &	$-$77 	&2	&$-$12\\
5683.4709 & 4.742 & 0.011 & $-$0.031 & SOPHIE &	60 	&8	&$-$9\\
5685.4986 & 4.620 & 0.010 & $-$0.025 & SOPHIE &	$-$62 	&14	&2\\
5687.4977 & 4.734 & 0.010 & $-$0.037 & SOPHIE &	52 	&3	&5\\
\bottomrule
\end{tabular}
\label{W56_RVtable}
\end{table}

\begin{table}
\caption{Radial velocity and line bisector span measurements of WASP-57.} 
%\begin{tabular*}{0.5\textwidth}{@{\extracolsep{\fill}}lccrl} 
\begin{tabular}{cccccrcr} 
\toprule \toprule
BJD & RV & $\sigma_{\rm RV}$ & V$_{span}$& Instrument & ${\rm RV}-\gamma$ & V$_{span}-<$V$_{span}$$>$&O -- C\\
$-$2\,450\,000 &(\kms)&(\kms)&(\kms)&& (m\,s$^{-1}$)&(m\,s$^{-1}$)&(m\,s${^{-1}}$)\\
\midrule
5646.5086 & -23.143 & 0.018 &  -0.015 &  SOPHIE	&71	& $-$17	&0\\
5647.5231 & -23.255 & 0.073 &  -0.043 &  SOPHIE	&$-$41	& $-$45	&58\\
5661.5093 & -23.278 & 0.015 &  -0.021 &  SOPHIE	&$-$64	& $-$23	&19\\
5662.5905 & -23.182 & 0.014 &  -0.003 &  SOPHIE	&32	&  $-$5	&7\\
5668.4671 & -23.137 & 0.042 &   0.102 &  SOPHIE	&77	&   100	&13\\
5670.6029 & -23.276 & 0.037 &   0.110 &  SOPHIE	&$-$62 	&   108	&13\\
5671.5218 & -23.132 & 0.043 &  -0.022 &  SOPHIE	&82	& $-$24	&$-$10\\
5672.6144 & -23.258 & 0.024 &  -0.036 &  SOPHIE	&$-$44 	& $-$38	&$-$2\\
5681.4680 & -23.333 & 0.014 &   0.010 &  SOPHIE	&$-$119 &     8	&$-$27\\
5686.4934 & -23.192 & 0.015 &  -0.062 &  SOPHIE	&22 	& $-$64	&$-$4\\
\hline
5627.7898 & -23.304 & 0.028 &   0.094 &  CORALIE&$-$90 	&   116	&7\\
5648.7304 & -23.136 & 0.022 &  -0.034 &  CORALIE&78	& $-$12	&$-$6\\
5679.8603 & -23.169 & 0.028 &  -0.043 &  CORALIE&45	& $-$21	&$-$25\\
5680.7396 & -23.135 & 0.027 &  -0.137 &  CORALIE&79	&$-$115	&38\\
5683.7970 & -23.190 & 0.025 &  -0.088 &  CORALIE&24	& $-$66	&29\\
5684.6299 & -23.333 & 0.025 &  -0.122 &  CORALIE&$-$119	&$-$100	&$-$25\\
5685.7140 & -23.139 & 0.020 &  -0.017 &  CORALIE&75	&     5	&$-$17\\
5689.8100 & -23.278 & 0.031 &  -0.012 &  CORALIE&$-$64	&    10	&8\\
5692.8069 & -23.227 & 0.047 &  -0.005 &  CORALIE&$-$13	&    17	&78\\
5705.6193 & -23.125 & 0.042 &   0.014 &  CORALIE&89	&    36	&$-$6\\
5722.5786 & -23.116 & 0.024 &   0.065 &  CORALIE&98	&    86	&10\\
5763.5386 & -23.280 & 0.023 &  -0.084 &  CORALIE&$-$66	& $-$62	&$-$9\\
5764.5606 & -23.247 & 0.025 &  -0.006 &  CORALIE&$-$33	&    16	&$-$7\\
5765.5833 & -23.119 & 0.028 &  -0.000 &  CORALIE&95	&    22	&5\\
5767.5688 & -23.201 & 0.035 &   0.047 &  CORALIE&13	&    69	&2\\
\bottomrule
\end{tabular}
\label{W57_RVtable}
\end{table}

\begin{table}
\caption{Theoretical evolutionary models for WASP-54, WASP-56 and WASP-57.} 
%\begin{tabular*}{0.5\textwidth}{@{\extracolsep{\fill}}lccrl} 
\begin{tabular}{|c|cc|cc|cc|cc|} 
\toprule
Model&\multicolumn{2}{c|}{Padova} & \multicolumn{2}{c|}{YY}&\multicolumn{2}{c|}{Teramo}&\multicolumn{2}{c|}{VRSS}\\
& \mstar~(\msun)&Age (Gyr)&\mstar~(\msun)&Age (Gyr)&\mstar~(\msun)&Age (Gyr)&\mstar~(\msun)&Age (Gyr)\\
%& \tinymsun&\tinyGyr&\tinymsun&\tinyGyr&\tinymsun&\tinyGyr&\tinymsun&\tinyGyr\\
%\cmidrule(c){1-9}
\midrule
%\multicolumn{3}{|c|}{Schedulers} \\ \hline
%\hline 
\noalign{\smallskip}\hline
WASP-54& 1.1$^{+0.1}_{-1.0}$&6.3$^{+1.6}_{-2.4}$&1.08$^{+0.09}_{-0.02}$&6.95$^{+0.96}_{-1.86}$&1.08$^{+0.09}_{-0.09}$&6.0$^{+1.5}_{-0.7}$&1.10$^{+0.04}_{-0.05}$& 5.8$^{+1.2}_{-0.7}$\\
\noalign{\smallskip}\hline%\hline
WASP-56&$0.96\pm0.04$&7.6$^{+3.7}_{-3.5}$&1.01$^{+0.03}_{-0.04}$&6.2$^{+3.1}_{-2.1}$&0.97$^{+2.93}_{-0.04}$&9.7$^{+3.6}_{-3.7}$&0.95$^{+0.04}_{-0.05}$&$9.7\pm3.7$\\
\noalign{\smallskip}\hline%\hline 
WASP-57&0.92$^{+0.02}_{-0.06}$&0.88$^{+4.53}_{-0.71}$&$0.89^{+0.04}_{-0.03}$&2.6$^{+2.2}_{-1.8}$&0.87$^{+0.04}_{-0.05}$&3.5$^{+3.5}_{-2.2}$&$0.90\pm0.04$&1.8$^{+3.5}_{-1.6}$\\
\noalign{\smallskip}\hline%\hline
\bottomrule
\end{tabular}
\label{tracks}
\end{table} 
\twocolumn

\end{document}